\documentclass[preprint]{aastex}
\usepackage{amssymb,amsmath,epsfig}

\newenvironment{boxedequation*}[1]{\begin{equation*}
\boxed{#1}}{\end{equation*}}

\newcommand{\mylabelf}[1]{\label{fig_#1}}

\newcommand{\mylabels}[1]{\label{sec_#1}}
\newcommand{\mylabelt}[1]{\label{tab_#1}}

\newcommand{\fig}[1]{Figure~\ref{fig_#1}}

\newcommand{\sect}[1]{Section~\ref{sec_#1}}
\newcommand{\tabl}[1]{Table~\ref{tab_#1}}

\shorttitle{Fiber Scrambling at Lick and Keck} 
\shortauthors{J. Spronck et al}

\begin{document}



\title{Fiber scrambling for high-resolution spectrographs. I. Lick Observatory}
\author{Julien F.P. Spronck, Debra A. Fischer, Zachary A. Kaplan, Christian Schwab and Andrew Szymkowiak}
\affil{Yale University}
\affil{New Haven CT, USA}
\email{julien.spronck@yale.edu, julspronck@gmail.com}

\begin{abstract}
In this paper, we report all results obtained with a fiber scrambler on the Hamilton spectrograph at Lick Observatory. We demonstrate an improvement in the stability of the instrumental profile using this fiber scrambler. 
Additionally, we present data obtained with a double scrambler that further improves the stability of the instrument by a factor 2. These results show that errors related to the coupling between the telescope and the spectrograph are the dominant source of instrumental profile variability at Lick Observatory. In particular, we show a strong correlation between instrumental profile variations and hour angle, most likely due to pointing-dependent illumination of the spectrograph optics.
\end{abstract}


\section{Introduction}
\mylabels{intro}

Since the discovery of the first exoplanet \citep{Mayor(1995)}, more than 650 planets have been confirmed, among which more than 400 have been found using the radial velocity method \citep{Wright(2011),Schneider(2011)}.
Currently, state-of-the-art spectrometers, such as HARPS \citep{Mayor(2003)} on the 3.6-m telescope in La Silla and HIRES on Keck I \citep{Vogt(1994)}, typically achieve precisions of 1-3~m~s$^{-1}$ \citep{Mayor(2008),Howard(2010)}.
This only permits the detection of planets with RV amplitudes comparable to or slightly less than the measurement errors, typically Super-Earths or Neptune-mass planets in relatively short period orbits, or more massive Jupiter-like planets out to several AU.
The Earth induces a reflex velocity of only 9~cm~s$^{-1}$ on the Sun, so the search for true Earth analogs requires Doppler precisions of about 10-20~cm~s$^{-1}$, corresponding to typical spectral line shifts across one ten-thousandth of a pixel. Further complicating the analysis, the periodicity of this shift occurs over time scales of months or year for the most interesting planets in the so-called habitable zone. This requirement for a measurement precision of 10~cm~s$^{-1}$ leads to the demand for an instrument that exceeds the stability of current instruments.

In order to reach the desired precision, we must reduce variability in the instrumental profile, which cross-talks with the modeled wavelength parameters and introduces errors in our measurement of the Doppler shift. In many spectrographs, the starlight is coupled from the telescope to the instrument using a narrow slit. However, the slit illumination varies rapidly because of changes in seeing, focus and guiding errors. 
Changes in slit illumination affect the spectrum in two ways. 
Since the spectral lines are direct images of the slit, changes in slit illumination produce changes in the shape of the spectral lines. Additionally, variations in slit illumination can result in changes in the illumination of the spectrograph optics. This will in turn introduce different aberrations, which will change the instrumental response.
Mathematically, these two effects are modeled simultaneously by convolving the spectrum with the instrumental profile. Any instrumental profile variability impedes our ability to recover Doppler shifts with the desired precision. If the instrumental profile were unchanging, variations in the final extracted spectrum would be dramatically reduced. Thus, instrumental profile stability has become a focus of current instrumentation work. 


Optical fibers provide an excellent way to reduce variability in the illumination of the spectrograph. Fibers have been used since the 1980's to couple telescopes to high-precision spectrographs \citep{Heacox(1992)}. The throughput of fibers was initially low, however, they offered unprecedented convenience in mechanical design. The attribute of fibers that is particularly important today for high-precision Doppler measurements is the natural ability of optical fibers to scramble light \citep{Heacox(1980),Heacox(1986),Heacox(1988),Barden(1981)}, that is to produce a more uniform and constant output beam, independently of the input. Because light from the telescope must be efficiently coupled into the fiber, the fiber diameters must match the size of the seeing disk (generally 100 microns or more), so multi-mode fibers are required.

We have designed an optical fiber feed, FINDS\footnote{The installation of the fiber feed, named FINDS as in FINDS Exo-Earths, was made possible by the support of the Planetary Society.}, for the Hamilton spectrograph at Lick Observatory \citep{Spronck_SPIE(2010),Spronck_Book(2011)} that stabilizes the illumination of the spectrograph and the instrumental profile. In this paper, we report the different tests carried out with the fiber scrambler on the Hamilton spectrograph at Lick Observatory \citep{Vogt(1987)}. In \sect{terms}, we define the terminology used throughout this paper. In \sect{design}, we present the design of the Lick fiber scrambler. In \sect{guiding}, we report some guiding tests that we have performed with and without fiber feed. In \sect{pupil} and \sect{psf}, we analyze the intensity distribution evolution after the slit in the far-field and in the near-field. In \sect{slitfiber}, we show a comparison between slit and fiber instrumental stability using the Hamilton spectrograph. In \sect{doublescrambler}, we present the results obtained with a double scrambler at Lick Observatory. In \sect{psfvar}, we present evidence of instrumental profile variations as a function of hour angle. Finally, radial velocities are presented in \sect{rv}.

\section{PSF and other misused acronyms}
\mylabels{terms}

Historically in high-resolution spectrometers, the terms point spread function (PSF) and 
line spread function (LSF) have been used interchangeably by astronomers to describe
the way an infinitely narrow spectral line is broadened by the instrument. 
This broadening occurs either by choice of a finite size slit or fiber that sets the spectrograph resolution or
by the quality of the optics and the alignment of the spectrograph that degrades the resolution.
The broadened line for any given wavelength is simply the image of the slit or the fiber by the optical system that is the spectrograph.
Neither PSF nor LSF should be used in this sense, since they already have a meaning that differs from this.

PSF is a term that should strictly be used in optical design: as its name indicates, it is the response of a given optical system to a point (dimensionless) source of light.
The slits and multi-mode fibers used in all current spectrographs are not point sources. They have a finite size. 
If they could be considered as point sources, doubling their size should not affect the spectrograph resolution.
The PSF is a characteristic of the instrument and only of the instrument, not of the slit illumination.
It only tells us about the quality of the optical system and is independent on the choice of slit.
In good approximation, the actual image on the CCD is the convolution of the PSF by the slit illumination function (the light intensity distribution over the slit, for which a good approximation is 0 outside the slit and 1 within the slit).

Similarly, the term LSF is used to describe the response of the instrument to a line source, i.e. an infinitely narrow and infinitely long slit.
Again, our slits are finite and the image of the slit on the CCD is the convolution of the LSF by the slit illumination function.

A more accurate term would be spectral line spread function (SLSF), which as its name indicates, simply represents the broadening and distortion of a spectral line by the instrument:
this includes both PSF (or LSF) and the slit illumination function. Therefore, throughout this paper, we will only refer to SLSF when we mean broadening and distortion of spectral lines.

\section{Design}
\mylabels{design}

The Hamilton is an echelle spectrograph with a resolution ranging from 60,000 to 100,000. 
It is permanently mounted at the Coud\'{e} focus of the Shane 3-m telescope.
Alternatively, the spectrograph can be fed by a 0.6-m Coud\'{e} Auxiliary Telescope (CAT) when the 3-m is not configured for Coud\'{e} observing.
The Hamilton covers a spectral range between 350 and 900~nm. 
To obtain precise radial velocities with this spectrograph, we use an iodine cell as a wavelength calibrator \citep{Butler(1996)}.
As starlight passes through the cell, the molecular iodine imposes thousands of absorption lines in the stellar spectrum.
The iodine lines are mostly present between 500 and 600~nm. Therefore, all results presented here correspond to this limited spectral range.

Light from the telescope falls onto the slit with a focal ratio of F/36. 
The slit used for our regular observing program is 640~$\mu$m wide, which corresponds to 1.2~arcsec on sky. Typical seeing varies from 1.5 to 2~arcsec. 
All slits are oriented along the meridian and atmospheric dispersion is not compensated.
Light reflected off the edges of the slit is then directed towards the guider camera. 
A green filter is used for guiding since we are mostly interested in the spectral range covered by the iodine lines.

A schematic drawing of the fiber feed is depicted in \fig{setup}. 
We have chosen to position this fiber system after the slit in order to ease the integration and keep the current acquisition and guiding system. 
Light coming from the slit is picked off by a mirror $M_1$ (which can move in and out of the way) and is focused 
with an achromatic doublet $L_1$ onto the optical fiber. $L_1$ transforms the F/36 beam into a F/4 beam before entering the fiber. 
For alignment purposes, a pellicle beam splitter is inserted into the beam between $L_1$ and the fiber. 
Light reflected off the fiber front surface would then be reflected by the pellicle and sent to a camera (Basler SCA640). 
This pellicle and camera are only used to check that the fiber and the slit are co-aligned. 
The pellicle would then be removed during observations.
After the fiber, a second achromatic doublet $L_2$ is used to give the beam its original
focal ratio and to re-image the end of the fiber onto a virtual source located at the same distance from the collimator as the original slit.
Light is then sent to the collimator with another mirror $M_2$. The fiber used in these tests is a 15-m FBP100120140 from Polymicro. Both ends are anti-reflection coated ($R < 0.5\%$ for 500-600~nm).

\begin{figure}[htbp]
	\plotone{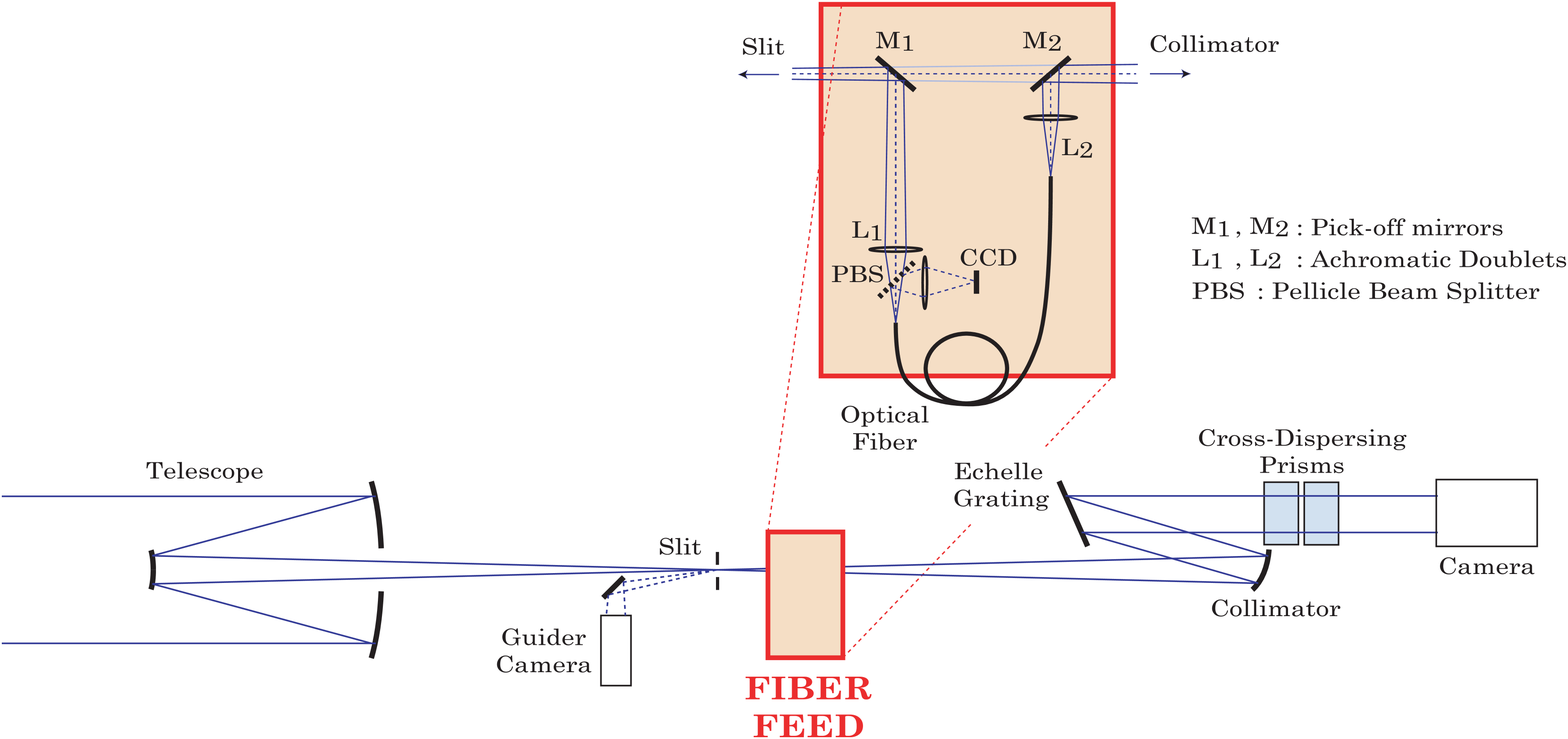}
	\caption{Schematic drawing of the Lick fiber feed.}
	\mylabelf{setup}
\end{figure}

After setting the fiber feed behind the slit, the throughput was 65$\%$\footnote{The definition of throughput that we used here is the decrease in number of counts on the final CCD with or without fiber feed. 
A 65$\%$ throughput means that the exposure time without fiber should be 0.65 times the exposure time with fiber to achieve the same SNR.}. 
The throughput is rather low for such a short fiber feed. We believe this low throughput may have been caused by the focal ratio degradation of the fiber and by non-optimized optical design. 
However, for the purpose of these tests, a high throughput was not essential and therefore was not optimized in this regard.

The fiber has a 1.6~arcsec field of view (100-$\mu$m fiber fed at F/4), which is larger than the slit used in our regular observing program (1.2~arcsec). 
However, because the width of a round fiber varies as a function of height from 0 to its diameter, the effective width of the fiber is smaller than its diameter. 
This is referred to as the resolution boost factor of fiber-fed instruments \citep{Vaughnn(1994)}. This factor is $\approx 1.18$ for round fibers compared to a slit.
This gives an effective width of 1.3~arcsec, which yields a $\approx 10$\% loss in resolution compared to our regular slit.

\section{Guiding tests}
\mylabels{guiding}

As part of the testing phase at Lick, we carried out a simple guiding test with the fiber on the 3-m Shane telescope. 
In this test, we obtained spectra of the iodine cell 
illuminated by a bright B-type star. We first obtained spectra using the slit without 
the fiber feed (i.e., typical of our usual program observations). We then inserted the fiber scrambler in the light path for the sets of test observations. For each set, we intentionally guided the telescope with the image of the star (a) 
approximately 1~arcsec on one side of the slit center, (b) centered on the slit and (c) 1~arcsec on the other side of the 
slit center. We used the 640-$\mu$m slit for all observations (even for fiber observations, since the fiber is located after the slit). This slit corresponds to 1.2~arcsec on sky (and a two-pixel resolution of 60,000).
Typical seeing varies from 1.5 to 2~arcsec and guiding errors of 0.2-0.3~arcsec typically occur. 
At Lick, there is no atmospheric dispersion compensator and the slit is oriented along the meridian. 
A green filter is used for guiding.

We cross-correlated each series of observations with spectra obtained when the star was properly guided at 
the center of the slit. \fig{guiding} shows the mean pixel shift as a function of the guiding position 
across the width of the slit for the central echelle order. The difference from this first test is remarkable: the iodine 
lines shift in the dispersion direction by up to 0.4 pixels ($\approx{0.02}$~\AA~at 600~nm) without a fiber. In contrast, we see that 
observations obtained with the fiber are far more stable with iodine line shifts of the order of 0.01 pixels ($\approx{0.0005}$~\AA~at 600~nm). 

The data points of \fig{guiding} correspond to cross-correlations of the 100 central pixels of the central echelle order. 
We repeated the same analysis for all echelle orders with iodine absorption lines (14) for slit (\fig{crosscororders}, left) and fiber (\fig{crosscororders}, right) observations.
\fig{crosscororders} (left) shows that the pixel shift introduced by guiding errors is wavelength-dependent for slit observations. 
This means that not only spectral lines are shifted on the CCD but also that the wavelength solution is affected by these guiding errors.
This wavelength-dependence in pixel shift introduced by guiding errors is probably due to the lack of atmospheric dispersion compensation and changes in the illumination of the spectrograph optics that introduce wavelength-dependent systematic errors.
We see on \fig{crosscororders} (right) that guiding errors induce almost no wavelength-dependent pixel shift when the fiber is used, due to its scrambling properties.

It is worth noting that, for absorption-based calibrators (such as iodine lines), the position of the lines and the wavelength solutions are not critical since wavelength calibration lines (i.e. the iodine lines) are imprinted in the spectrum.
However, even for iodine calibrated spectrographs, stabilizing the spectrum helps to reduce the variability in the SLSF and therefore reduces cross-talk between the SLSF and wavelength model in the Doppler analysis code.

\begin{figure}[htbp]
    \plotone{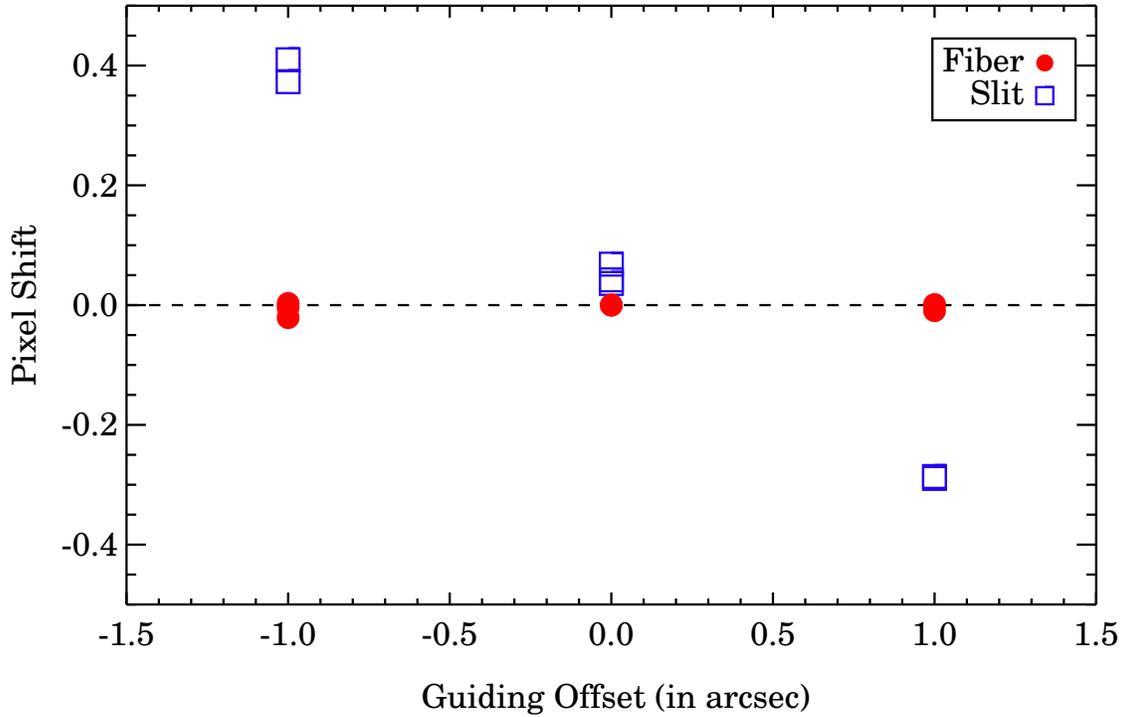}
	\caption{Mean pixel shift of the spectra (central echelle order) as a function of guiding. The blue rectangles 
correspond to the observations without fiber while red filled circles correspond to observations 
using the fiber feed.}
	\mylabelf{guiding}
\end{figure}

\begin{figure}[htbp]
	\plottwo{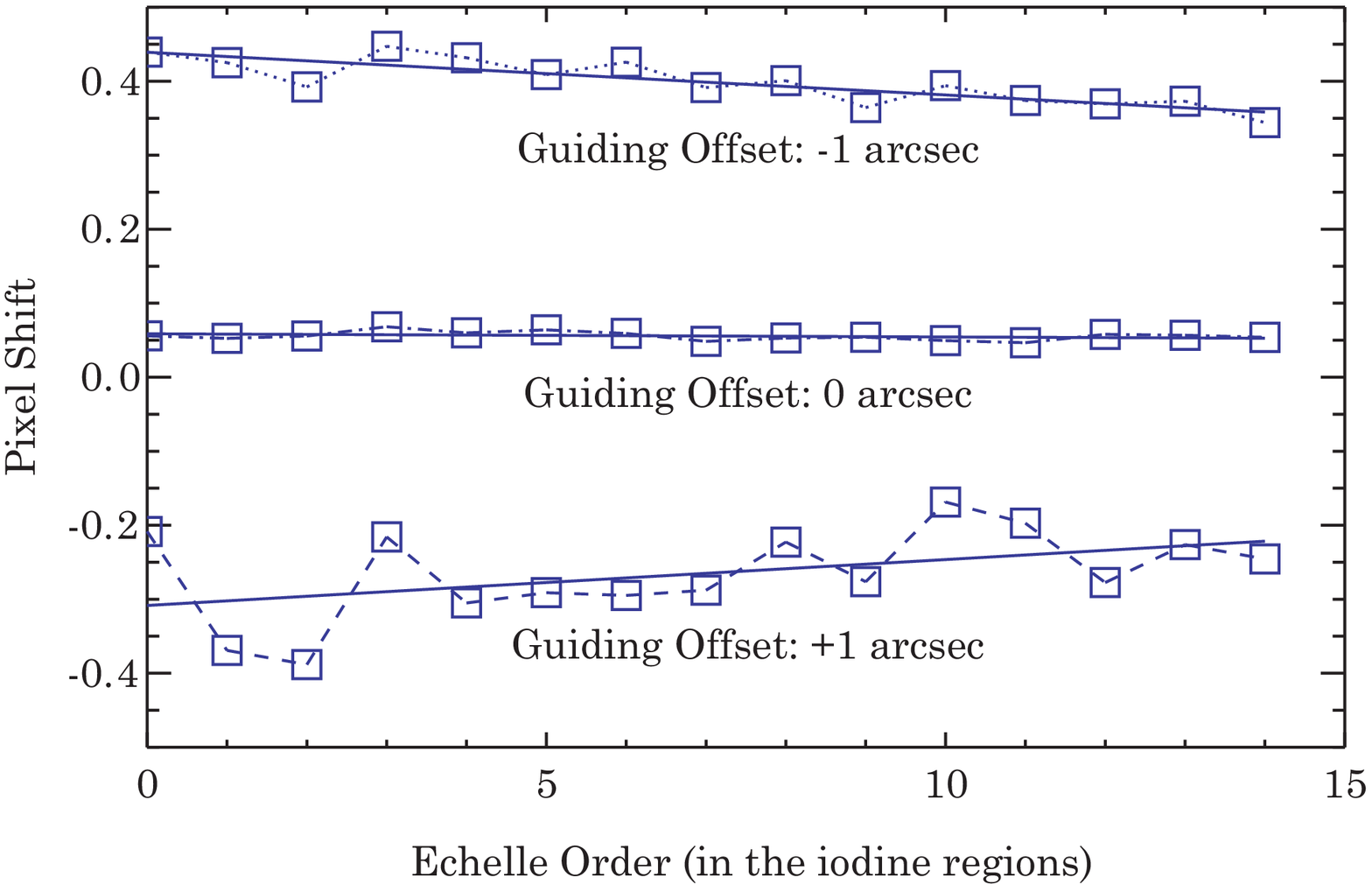}{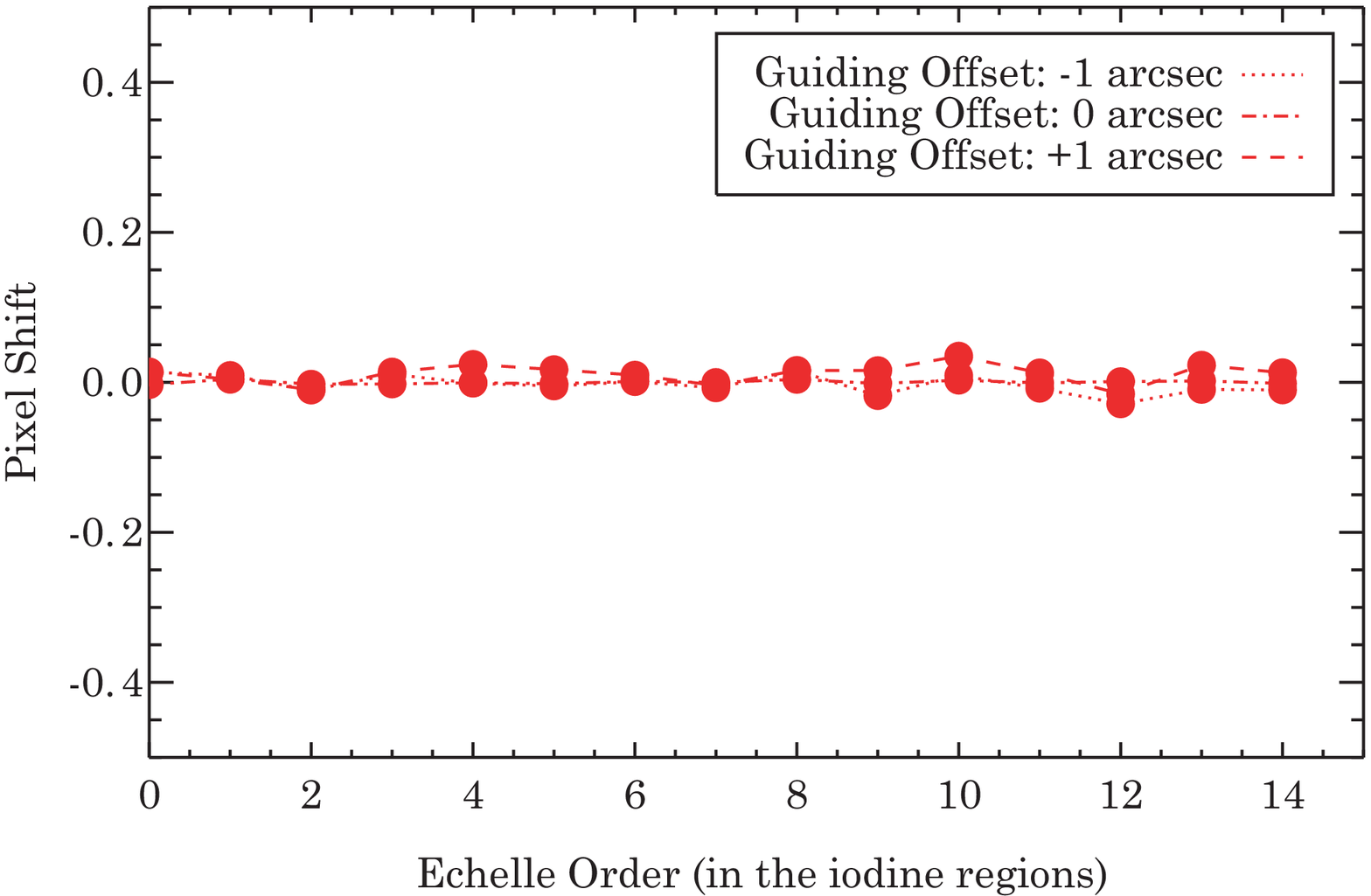}
	\caption{Pixel shift of the spectra as a function of guiding for all echelle orders in the iodine region for slit observations (left) and fiber observations (right).}
	\mylabelf{crosscororders}
\end{figure}

\section{Far-field measurements: pupil illumination}
\mylabels{pupil}

The second series of tests that we have performed consists in taking consecutive snapshots of the 
pupil illumination with and without the fiber to see the improvement in stability of the 
illumination of the spectrometer optics. 

As mentioned in \sect{design}, the Hamilton spectrograph can be fed by a 0.6-m Coud\'{e} Auxiliary Telescope (CAT) when the 3-m is not configured for Coud\'{e} observing.
For both telescopes, the guiding system and spectrograph control remain the same. 
However, the plate scale changes by a factor 5, which deteriorates the guiding performance when using the CAT.
For these measurements, we used the CAT guiding on the bright A-type star, Deneb. 

In order to measure the far-field after the fiber when guiding on a star, we used a lens and a commercial 752~x~480 pixel 8-bit CMOS camera (Orion StarShoot and Planetary Imaging Camera) that we positioned right after the fiber feed.
In this case, the lens is not used to image the fiber onto the camera but to fit the entire pupil on the detector.
\fig{pupil} includes three consecutive snapshots of the 
pupil illumination without (a) and with the fiber feed (b). The pictures were 
taken every 0.33 s without fiber and every 1 s with the fiber. 
Without fiber, we see a non-uniform intensity distribution and important variations from one picture to the other. The fiber 
dramatically improves the uniformity and the stability of the illumination of the spectrometer 
optics. Note that the brighter spot in the snapshots taken with the fiber comes from imperfections 
in the camera used to take the pictures. 
Even though \fig{pupil} illustrates the variability in pupil illumination when using the slit and shows the improvement brought by the fiber feed, the difference in pupil illumination when using the Shane and the CAT can be significant due to the different plate scales and the ratio between the aperture and the average size of turbulent cells.

\begin{figure}[htbp]
    \centering
	\begin{tabular}{ccc}
	\includegraphics[width=4.7cm]{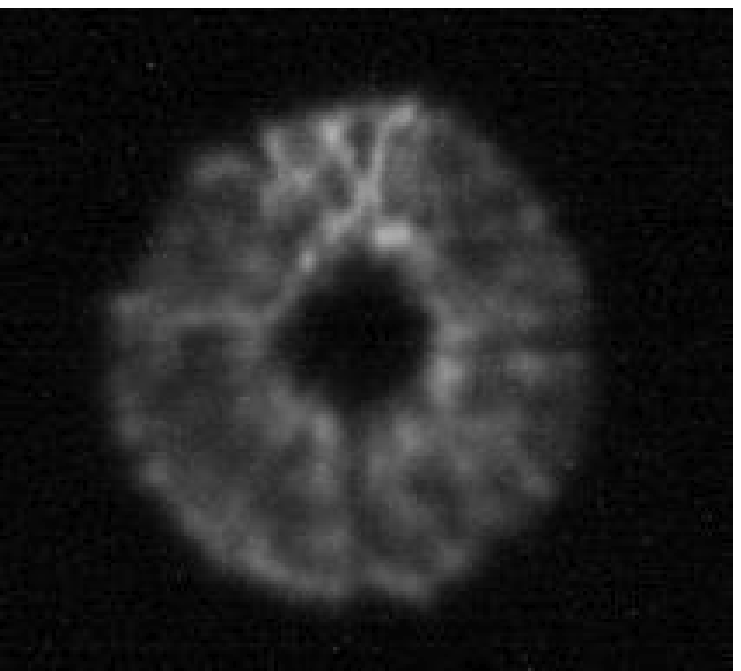}
	&\includegraphics[width=4.7cm]{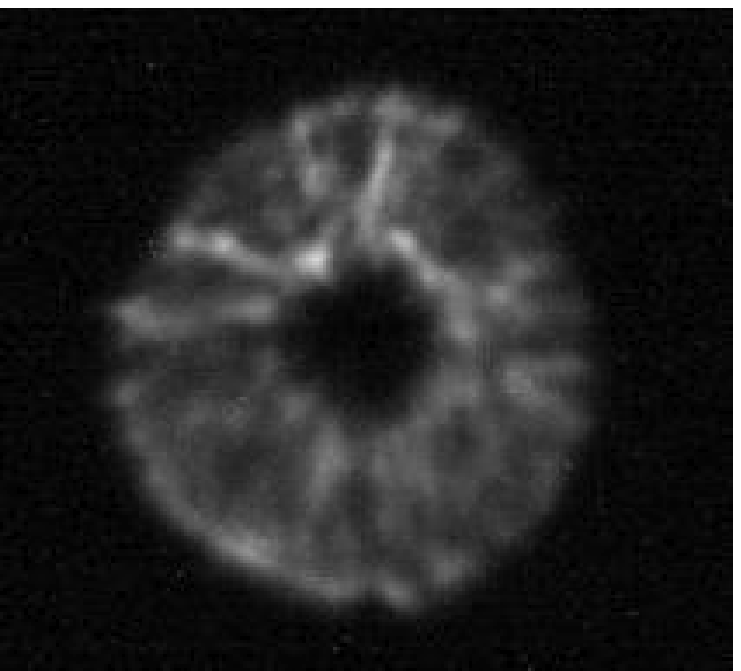}
	&\includegraphics[width=4.7cm]{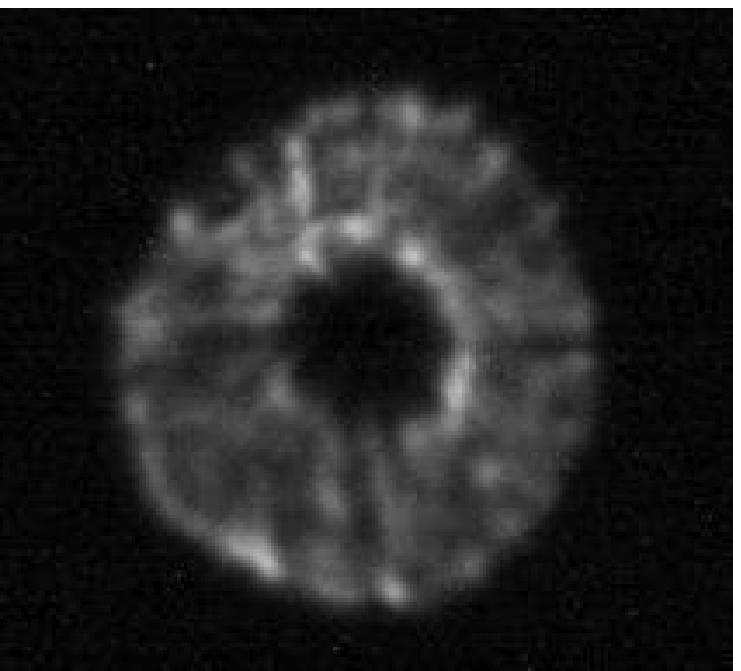}
	\\
	 &(a)& \\
	 & & \\
	\includegraphics[width=4.7cm]{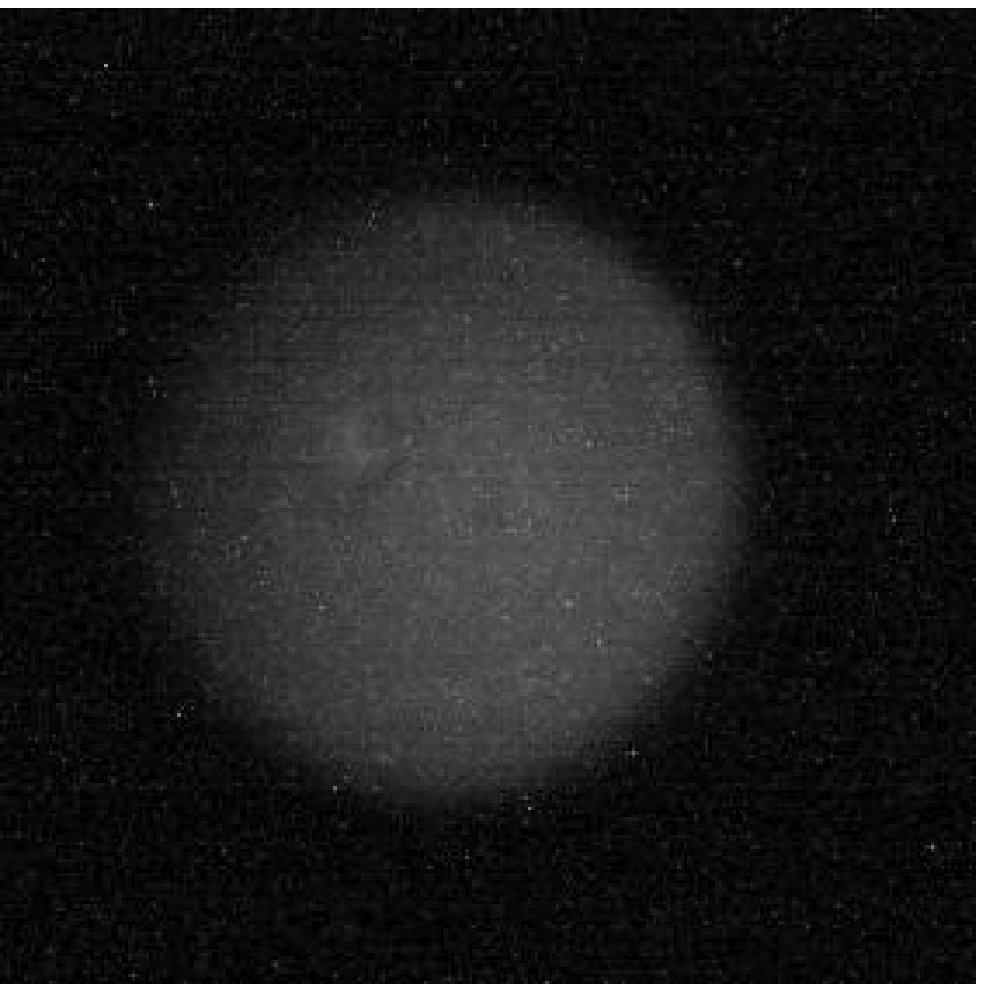}
	&\includegraphics[width=4.7cm]{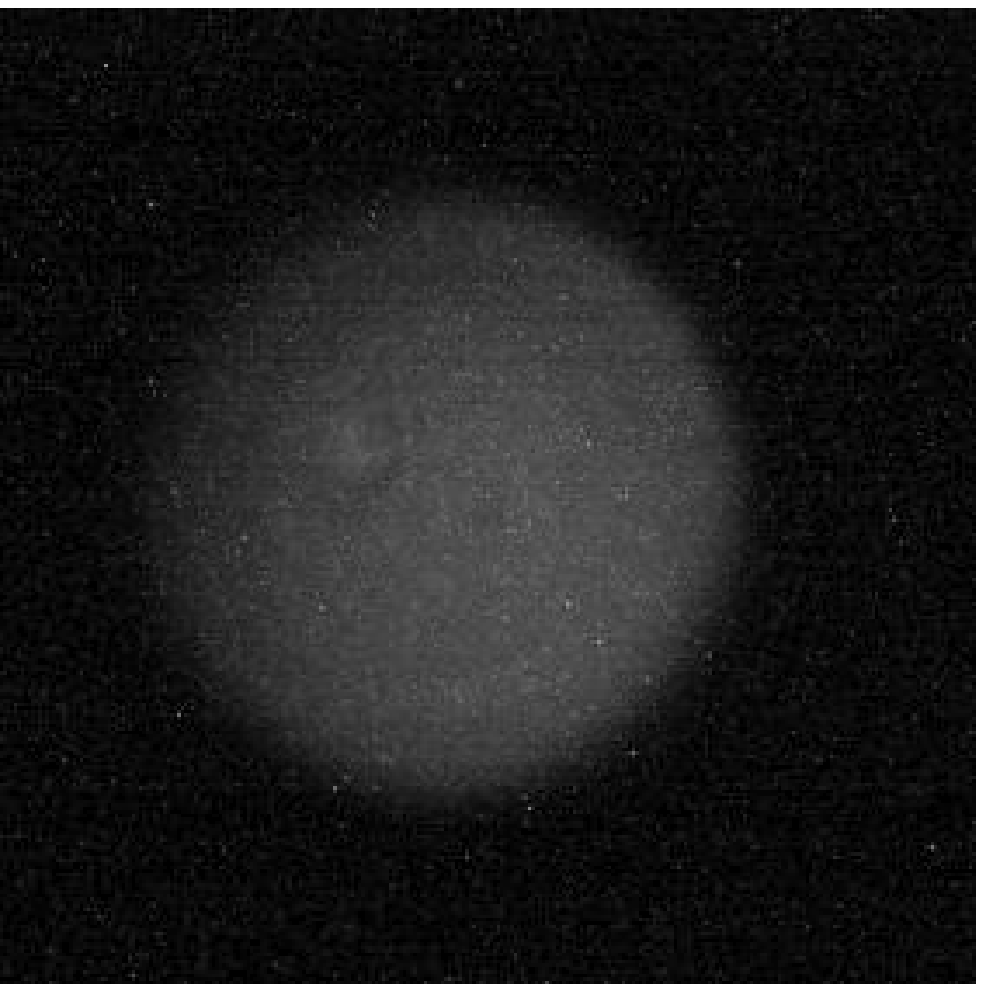}
	&\includegraphics[width=4.7cm]{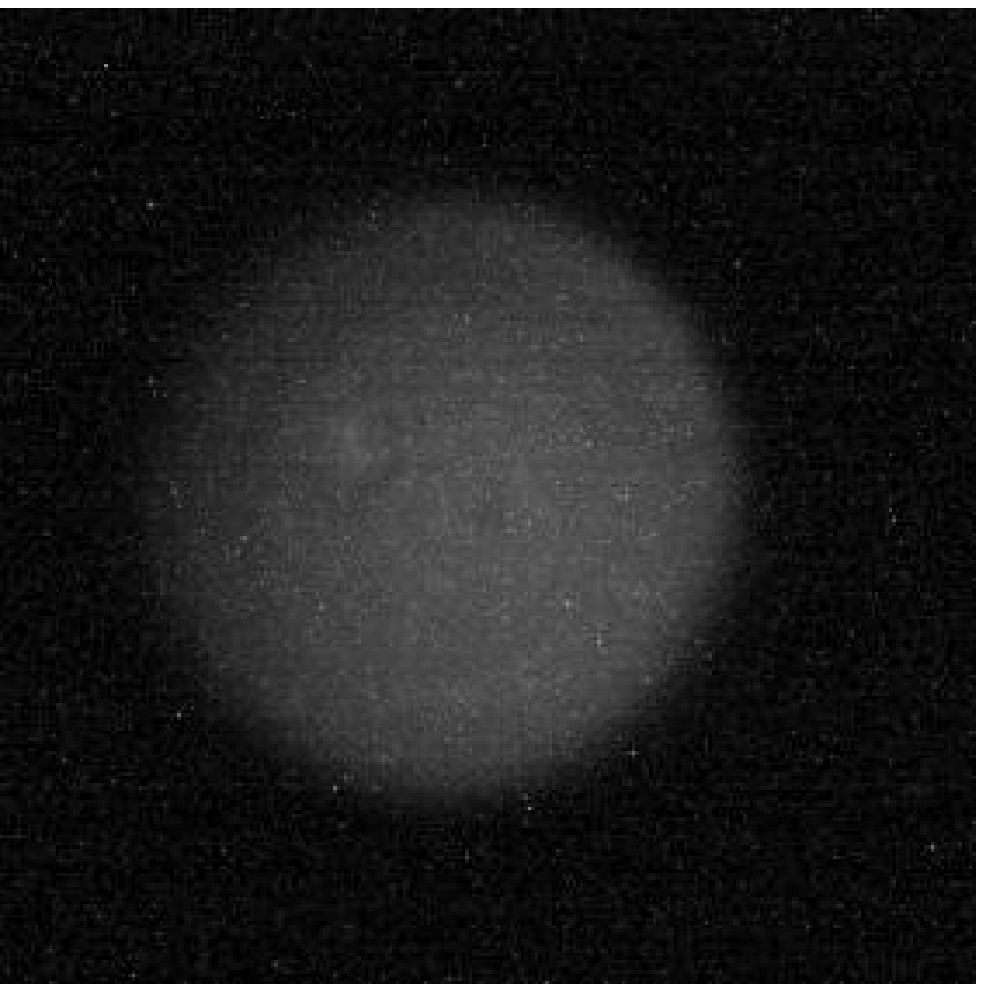}
	\\
	 &(b)& \\
	\end{tabular}
	\caption{Intensity distribution in the pupil plane without (a) and with (b) the fiber feed. These three consecutive snapshots span 1 sec (3~FPS) without the fiber and 3 sec (1~FPS) with the fiber.}
	\mylabelf{pupil}
\end{figure}

\section{Near-field measurements: SLSF Stability}
\mylabels{psf}

In order to analyze the SLSF stability, we pointed the 3-meter telescope at $\alpha$ Peg and we used the same lens and CMOS camera described in the previous section (\sect{pupil}). We focused the beam coming from either 
the fiber (when the fiber feed was used) or from the slit (when the fiber feed was not used) on the detector with a de-magnification of about 5. 
We then recorded the varying fiber or slit image on the detector as a function of time, which we refer to as near-field\footnote{The near-field pattern is the intensity distribution across the output face of the fiber. Commonly (but erroneously), the term near-field is used to describe the image of the output face of the fiber by an optical system. We will adopt this definition throughout this paper. Similarly, the term is also used for slit-fed spectrographs to describe the intensity distribution in the plane of the slit and the image of the slit by the optical system}. 
This near-field directly translates into SLSF since, for any given wavelength, the spectrograph merely takes an image of the output face of the fiber or of the slit.
Both data sets (with and without fiber) spanned 60 seconds of time. 
The fiber set was taken at 20~frames per second (FPS), and the slit data at 40~FPS. 
The two sets were taken back to back, therefore under similar seeing conditions.

Each image corresponds to a different time. We then mashed each image in the direction defined by the slit into a one-dimensional intensity distribution and performed a Gaussian fit to the 1-D SLSF.
We adopted three metrics to evaluate the stability of the SLSF through time (\fig{psf}): 
(a) the full-width at half maximum (FWHM) of the Gaussian, (b) the asymmetry of the Near-Field intensity pattern, which was estimated by calculating the displacement of the 25$\%$ bisector
compared to the center of the Gaussian and divided by the mean FWHM and (c) the position of the center of the Gaussian (also divided by the mean FWHM).
\fig{psf} depicts the comparison of these three SLSF metrics with and without fiber feed for a 60-second time span. 
The data were taken at 20 (fiber observations) and 40~FPS (slit observations). 
However, we binned all data into 0.5 second time bins in order to compare slit and fiber data.
For the three metrics (FWHM, asymmetry, position), the standard deviation of the binned time series is given with and without the fiber feed. 
These three quantities show improvements in stability by a factor 10 (10, 7 and 15 respectively).

All data presented in this section only spanned one minute of time. Taking long exposures should average down the fluctuations and should improve the stability. 
We checked the effect of binning into longer time bins (5 and 10 seconds) and we found very similar results. 
With 5-second (resp. 10-second) bins, the FWHM for fiber observations was 13 (resp. 19) times more stable than the FWHM obtained with the slit. 
Similar numbers were measured for the asymmetry and the centroid.

\begin{figure}[htbp]
    \centering
	\begin{tabular}{c}
	\includegraphics[width=8.5cm]{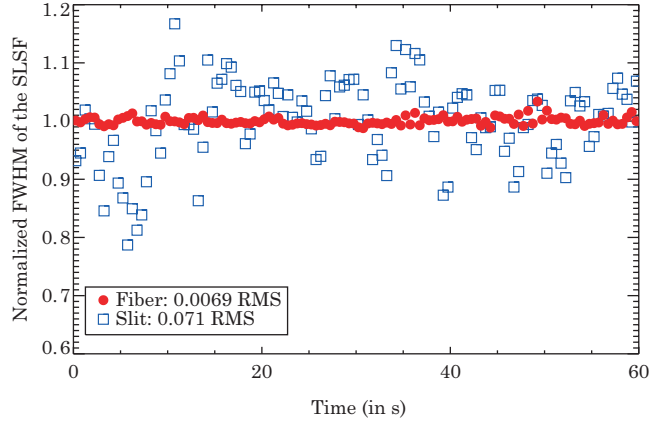}
	\\(a)\\
	\includegraphics[width=8.5cm]{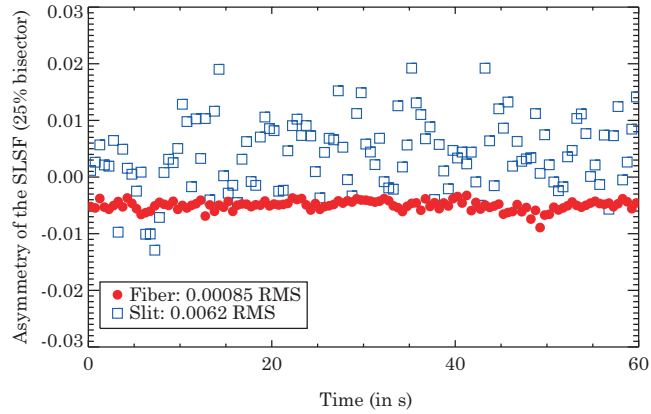}
	\\(b)\\
	\includegraphics[width=8.5cm]{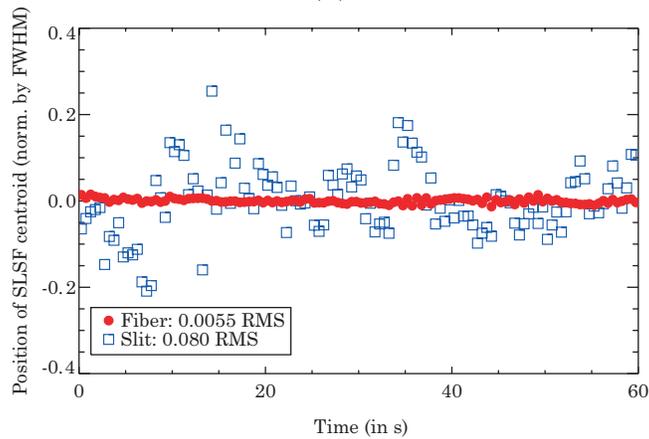}
	\\(c)\\
	 \\
	\end{tabular}
	\caption{Comparison of (a) FWHM, (b) asymmetry and (c) position of the center of the intensity pattern with (open squares) and without (filled circles) the fiber feed for a 60-second time span.}
	\mylabelf{psf}
\end{figure}

\section{Comparison between slit and fiber using the Hamilton spectrograph}
\mylabels{slitfiber}

In August 2010, tests were carried out to quantify the improvement in 
instrumental profile stability brought by the fiber scrambler and to identify the remaining sources of error. Observations of stars with known constant radial velocities were made on two consecutive nights. The weather and seeing conditions were nearly identical for both nights. The fiber scrambler was installed for the first night, and the regular observing slit ($640~\mu$m wide) was used on the second night. 

On both nights, an iodine cell was used. As starlight passes through the cell, the molecular iodine imposes thousands of absorption lines in the stellar spectrum. 
The SLSF is determined by modeling the narrow iodine lines in the spectrum \citep{Valenti(1995)}. 
We use a high resolution ($R\approx$~1,000,000) spectrum of our iodine cell with $\mbox{SNR}\approx 1000$, obtained with a Fourier Transform Spectrograph (FTS) at the EMSL division of the Pacific Northwest National Laboratory. 
This FTS iodine spectrum is convolved with the model SLSF description that gives a best fit to our observations. For both nights, the iodine cell was kept at a temperature of $50 \pm 0.2~C$. The line depths do not change measurably from observation to observation.

Our SLSF model consists of 17 Gaussians (1 central main Gaussian and 8 outer ones on each side). In this test, we are interested in the temporal SLSF stability, which can be studied by studying the FWHM of the SLSF, its asymmetry and its position as a function of time.

The SLSF is modeled for small (2~\AA) wavelength segments of the echelle spectrum to account for 2-D spatial variations (352 segments). However, for plotting simplicity, we will focus on the time variations occurring in one single wavelength segment. In reality, the time variation of the FWHM will also be spatially dependent. 
In other words, the SLSF corresponding to different wavelength segments will have different time behaviors.

In this section, we also adopted three metrics to evaluate the stability of the SLSF through time: 
(a) the full-width at half maximum (FWHM) of the SLSF, (b) the asymmetry of the SLSF, which was estimated by calculating the displacement between the SLSF centroid and the midpoint at the 90$\%$-level, divided by the mean FWHM and (c) the SLSF centroid (also divided by the mean FWHM).

\fig{slitfiber} depicts the evolution of these three metrics for the slit observations (blue squares) and for the fiber observations (red filled circles) through time. For both nights, the same sequence of observations were taken: a set of B stars, 50 observations of the star HD~161797, a second set of B stars, 50 observations of the star HD~188512 and a third set of B stars. HD~161791 and HD~188512 are slightly evolved subgiants that have been observed for several years at both Lick and Keck Observatories. HD~161797 has a significant linear trend and HD~188512 has a shallow linear trend. The residuals to these trends have an RMS at Keck and Lick of about 5 m/s, consistent with our measurement precision for similarly evolved stars. Given their long-term RV stability and brightness, these are ideal stars for making tests where many measurements must be obtained.

We observe trends in the SLSF FWHM, asymmetry and centroid for the slit observations as a function of time. This functional dependence on time for slit observations strongly suggests that the dominant factor in SLSF variation is the 
changing illumination of the slit due to monotonic changes in seeing or tracking 
through different hour angles (which induces systematic errors due to atmospheric dispersion and possibly star position-dependent optics illumination). 
In addition to the trends, we see two populations of data points for the slit observations, each corresponding to a different star. 
We will discuss these SLSF variations in more detail in \sect{psfvar}.

\fig{slitfiber} also shows improvement in instrumental profile stability when using the fiber 
scrambler (red solid dots). All metrics show an improvement by about a factor 2 between slit and fiber observations. However, there is still a slight linear (upward) trend in 
the fiber data (1\%-2\% PTV for the FWHM), indicative of incomplete scrambling with the fiber.

\begin{figure}[htbp]
    \centering
	\begin{tabular}{c}
	\includegraphics[width=8.5cm]{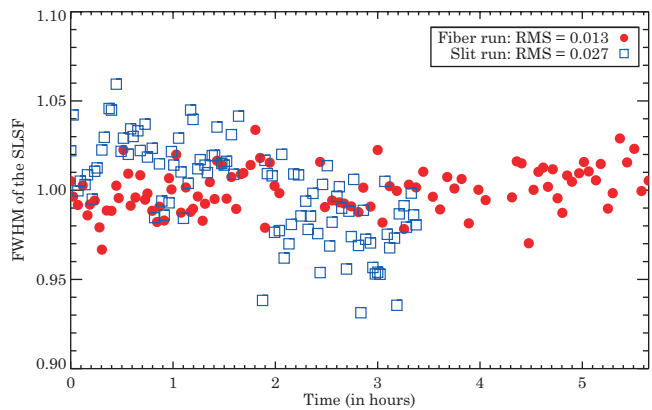}
	\\(a)\\
	\includegraphics[width=8.5cm]{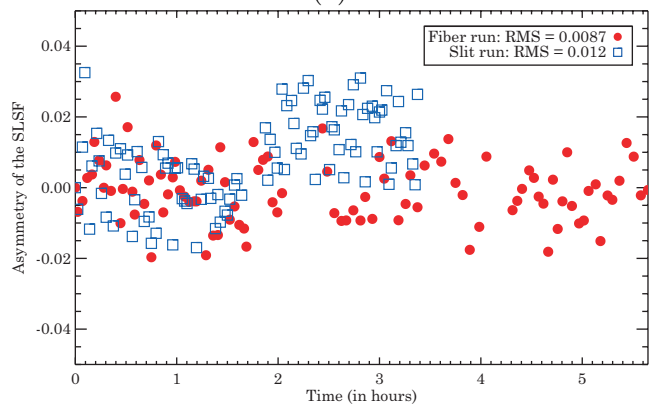}
	\\(b)\\
	\includegraphics[width=8.5cm]{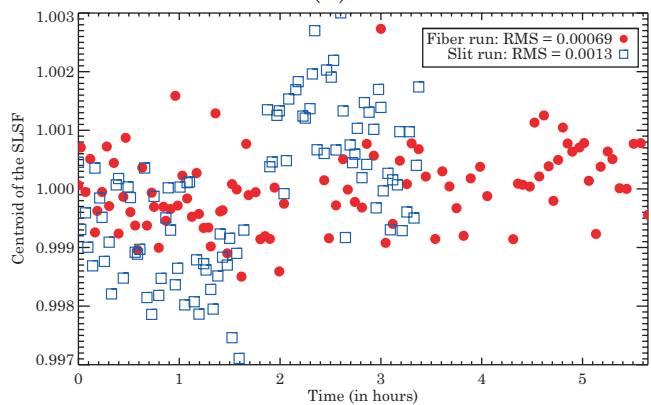}
	\\(c)\\
	 \\
	\end{tabular}
	\caption{Average FWHM of a Gaussian fit to the SLSF for all 
observations during Night 1 using the fiber (red filled circles) and Night 2 using the slit (blue squares).}
	\mylabelf{slitfiber}
\end{figure}

\section{Results with a double scrambler using the Hamilton spectrograph}
\mylabels{doublescrambler}

In August 2010, a double scrambler \citep{Hunter(1992),Avila(1998)} was designed and built (see \fig{doublescrambler}~(a)).
In this double scrambler, a ball-lens (1~mm in diameter, Edmund Optics) transforms the image of the fiber end in a pupil that is then injected into a second fiber.
The light from the second fiber is then sent to the spectrograph. As for previous tests, both fibers were 15-m FBP100120140 AR-coated fibers from Polymicro. Because of time constraints in the mechanical design phase, the double scrambler was not optimized 
and as a consequence, the throughput when used in the Hamilton spectrograph was rather low (15\% compared to the slit as opposed to 65\% with one fiber only).
This low throughput was mainly due to misalignments and the lack of mechanical adjustments.

The double scrambler test consisted in taking alternative sets of five B-star observations with the regular fiber scrambler (one fiber only) and with the double scrambler throughout the same night.
For each observation, we modeled the SLSF for all wavelength segments of the spectra. As in \sect{slitfiber}, for simplicity, we studied one single wavelength segment. In this segment, we calculated the FWHM and the asymmetry of the SLSF.

\fig{doublescrambler}~(b) depicts the evolution of the FWHM for the single fiber observations (red) and for the double scrambler observations (green) through the night, while \fig{doublescrambler}~(c) depicts the asymmetry as a function of time. Different symbols correspond to different sets of B stars and a linear fit is overplotted (dotted lines).
Even though the scale is different from \fig{slitfiber}, we see a linear trend in the fiber data (in red) in \fig{doublescrambler}~(b) and \fig{doublescrambler}~(c), indicating imperfect fiber scrambling. 
In the case of the FWHM, the amplitude of the variation is about 4-5\%. The SLSF obtained with the double scrambler is more stable throughout the night, with no significant (above errors) systematic trend. 

Instrumental noise can be broken down into two main components: errors due to coupling of the light to the instrument (varying fiber illumination due to guiding, tracking, seeing, focusing, lack of or incomplete atmospheric dispersion compensation, ...) and environmental instability (mechanical, temperature or pressure). 
The variability in the environment was quite similar for observations made with the slit, the round fiber and the double scrambler. However, relative to the slit observations, the SLSF stability improved significantly with the round fiber and then again with the double scrambler. This is an unambiguous demonstration that coupling errors are the dominant source of variability in the SLSF over short time scales (i.e., throughout a night).

Residual fluctuations from observation to observation in the double scrambler FWHM  have an amplitude of 1-2\%.
The source for these fluctuations has not yet been identified but possible culprits include modal noise in the fiber (since the fiber was not agitated to reduce modal noise), photon noise and modeling errors. 
We do not expect the environmental instability to be responsible for residual fluctuations because of the short time scale of the variability. 
If these fluctuations are not real, then our errors in modeling the SLSF represent a significant error source for precise radial velocity measurements.  
In particular, errors in the centroid of the SLSF will cross talk with the wavelength dispersion and introduce errors in our Doppler measurements.  
The trends depicted in \fig{doublescrambler}~(b) and \fig{doublescrambler}~(c) for the fiber without double scrambling are likely to produce systematic velocity errors that average out much more slowly than random errors.

\begin{figure}[htbp]
    \centering
	\begin{tabular}{c}
	\includegraphics[width=8.5cm]{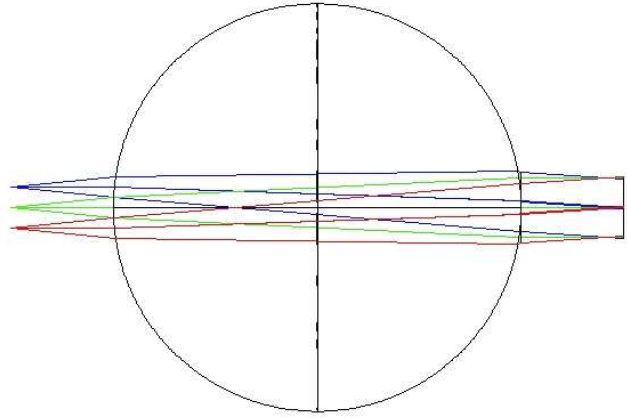}
	\\(a)\\
	\includegraphics[width=8.5cm]{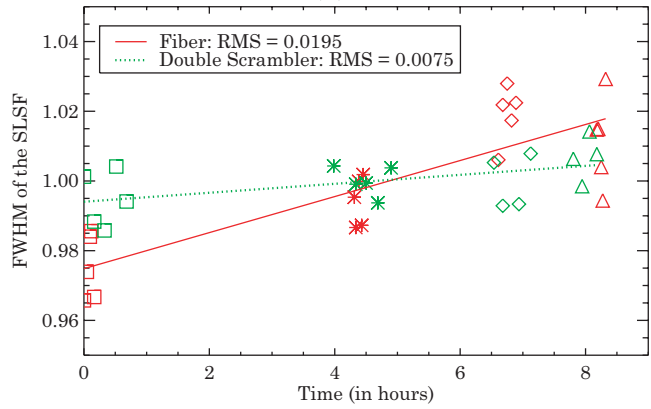}
	\\(b)\\
	\includegraphics[width=8.5cm]{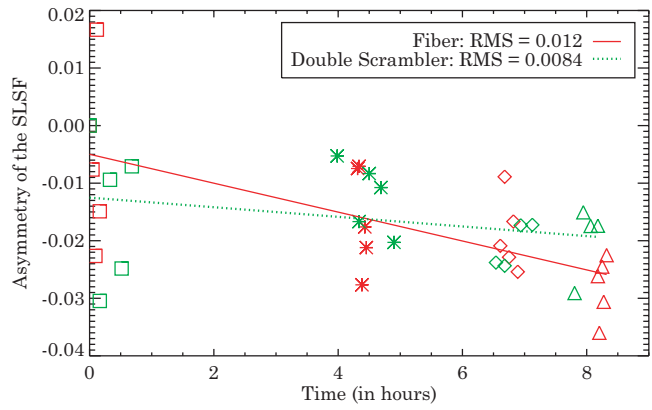}
	\\(c)\\
	 \\
	\end{tabular}
	\caption{(a) Schematic design of the double scrambler. (b) FWHM of the SLSF for B-star observations taken with the fiber (red) and with the double scrambler (green).
All observations were taken during the same night alternating between the fiber and with the double scrambler.}
	\mylabelf{doublescrambler}
\end{figure}

\section{On understanding SLSF variations}
\mylabels{psfvar}

As mentioned in \sect{slitfiber}, there seem to be evidences of SLSF variations as a function of the star position in the sky for slit observations. 
In order to demonstrate this, we have performed a test using the regular slit (1.2~arcsec on sky). We have chosen five B-type stars located at different positions in the sky. 
For each star, we have taken a set of three observations with iodine lines imprinted onto the stellar spectrum.
After going through the five stars, we went back to the first star and repeated the observations. 
In total, we did three cycles of three observations for each star (45 observations in total) during the same night.

Each spectrum is simply the convolution of the iodine lines with the SLSF, since these stars have no spectral absorption lines. 
Using an ultra high-resolution spectrum of the iodine cell and a Levenberg-Marquardt algorithm, 
we can model the SLSF for all observations for small wavelength segments of the echelle spectrum to account for 2-D spatial variations. 
We plotted in \fig{psfha}~(left) the FWHM of the SLSF for all 45 observations for one of the wavelength segments.

Each combination of color and symbol corresponds to a different star. 
The positions of the star were strategically chosen. Three stars have similar declinations between 0 and 3 degrees but different RA: 13h 35' (red asterisks), 14h 45' (yellow diamonds) and 17h 45' (cyan squares).
The two other stars had a similar RA of $\approx$ 15h 50' but different declinations: +15 degrees (green triangles) and -29 degrees (blue crosses).

\fig{psfha} shows the FWHM variations in slit observations for the repeated sets of our three stars as a function of time (left) and hour angle (right) for a given small wavelength segment.  If the SLSF varied as a function of time, we might hypothesize that the seeing was gradually changing and that this was correlated with a change in the FWHM parameterization of the SLSF.  However, the plot in \fig{psfha}~(left) does not show that effect. Furthermore, our measurements do not show evidence for a gradual change in the seeing. Instead, we see that variability in the SLSF for the slit observations is strongly correlated with hour angle (see \fig{psfha}~(right)), or with the position of the star in the sky. We can think of a few other physical causes for this: the variability could be correlated with airmass, atmospheric dispersion, or changing light path depending on the star position in the sky.

We do not believe that airmass is responsible for these SLSF variations. Consider the two stars corresponding to the cyan squares and the red asterisks. 
Both stars have the same declination but vary by four hours in right ascension. 
During the first set of observations, the red asterisks have $HA = 2$ and the cyan squares have $HA = -2$. Having the same declination, these stars at that time have the same airmass but a very different SLSF width.

If atmospheric dispersion was responsible for these variations, we would expect to see a color dependence in the SLSF variation. \fig{psfhaord} shows the FWHM of the SLSF as a function of hour angle for different physical parts of the CCD, i.e., as a function of wavelength. The top row shows the SLSF for three wavelength segments in one of the redder Echelle orders, the middle row shows segments from a green order in the iodine spectrum and the bottom row shows segments in one of the bluest orders of the iodine spectrum. So, top to bottom is the cross-dispersion direction and left to right is along the dispersion direction for segments of three Echelle orders.  Each of the nine segments will have a different SLSF because the quality of the optical system is field-dependent. If atmospheric dispersion is the underlying cause for the change in SLSF, then we would expect to see a difference in the response of the red, the green and the blue orders. However, looking from top to bottom, the trends are similar. Instead,  along the dispersion direction, we see systematic changes in the FWHM variations as a function of hour angle. Slit-fed spectrometers are unlikely to provide the extraordinary stability required in order to achieve extreme radial velocity precision.

The most likely explanation is that different star positions yield different light paths within the spectrograph. There could be a slight misalignment between the telescope and the spectrograph optical axes that introduces vignetting or that otherwise affects the dispersion of the spectrum as the telescope points away from the meridian.  The important point is that there can be subtle issues that introduce variability in the SLSF. Extraordinary stability will be required in order to achieve extremely high radial velocity precision. Slit-fed spectrometers are unlikely to provide stable illumination of the spectrograph optics and will therefore have a variable SLSF.

\begin{figure}[htbp]
	\plottwo{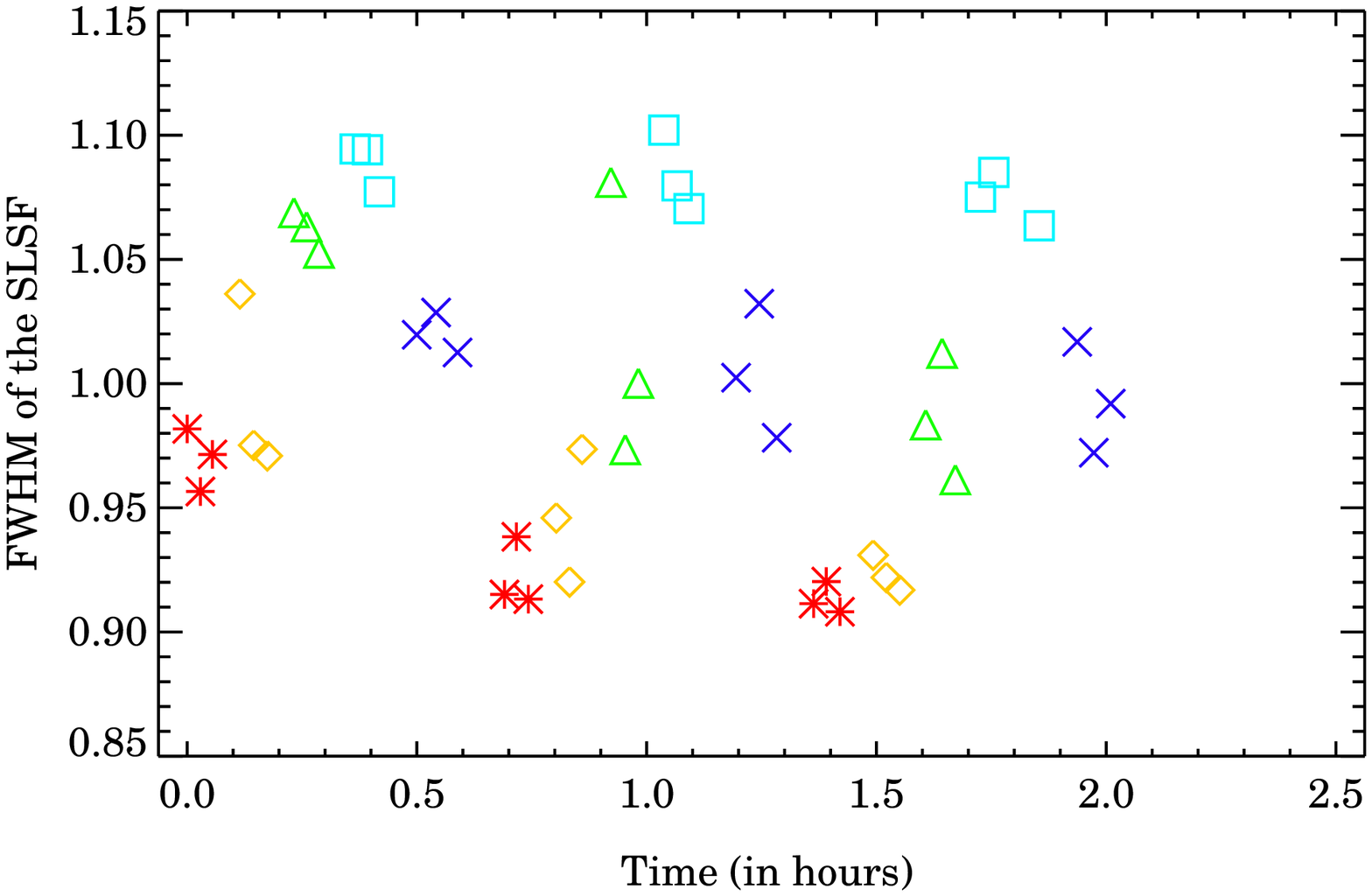}{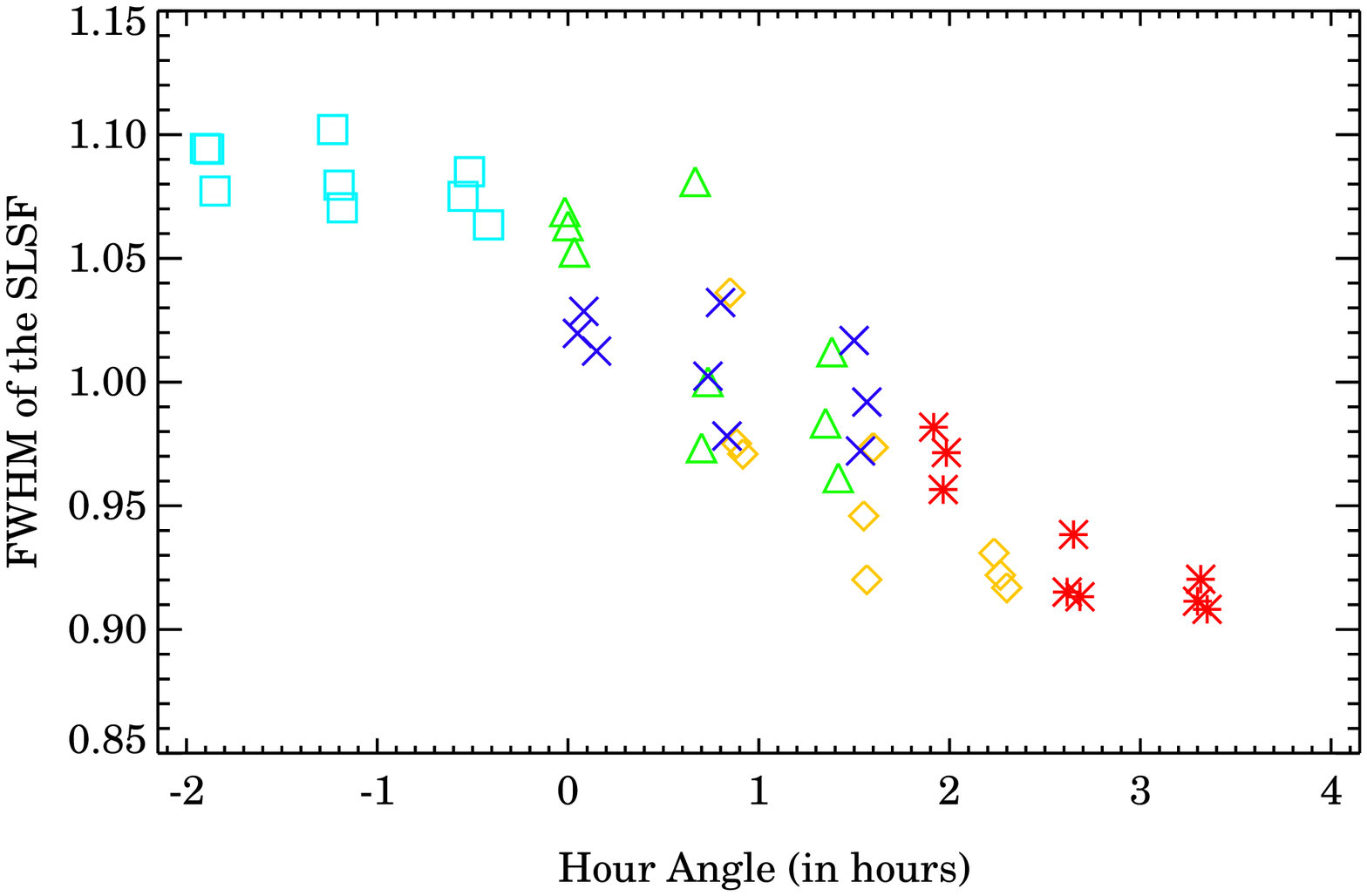}

	\caption{SLSF for all 45 B-star observations (left) as a function of time and (right) as a function of hour angle. 
	Each combination of color and symbol corresponds to a different star.}
	\mylabelf{psfha}
\end{figure}

\begin{figure}[htbp]
    \centering
	\begin{tabular}{ccc}
	\includegraphics[width=5cm]{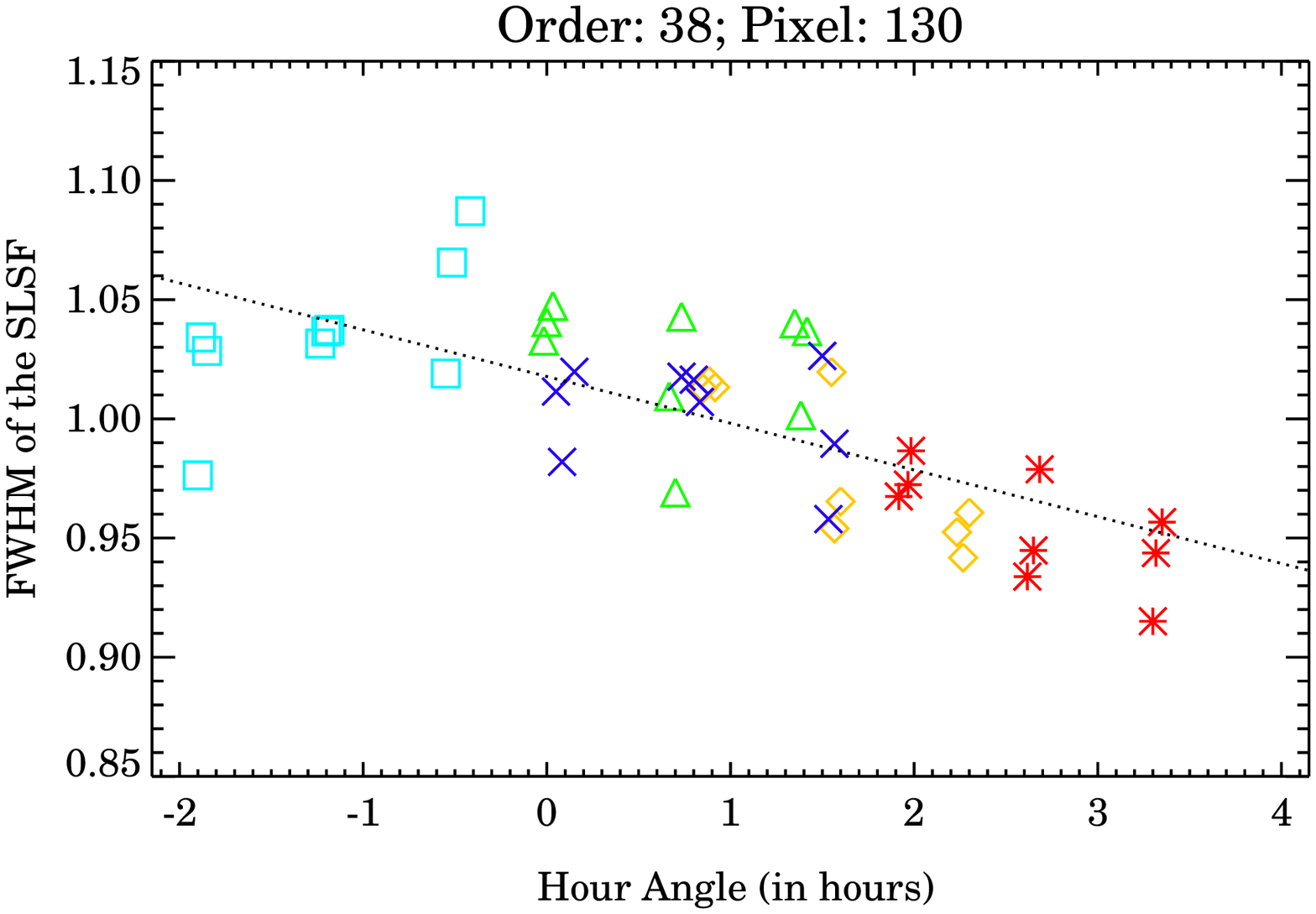}&\includegraphics[width=5cm]{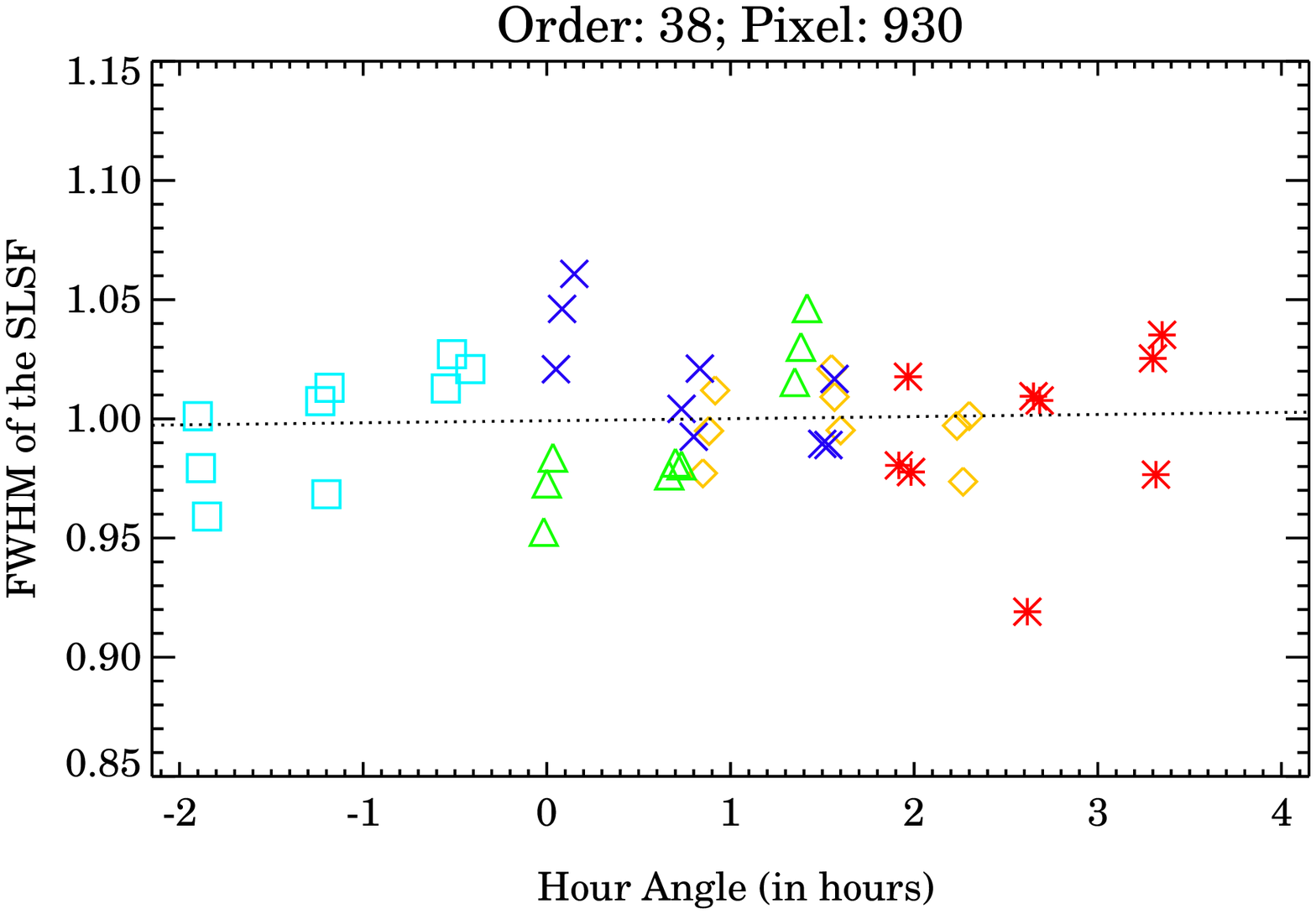}&\includegraphics[width=5cm]{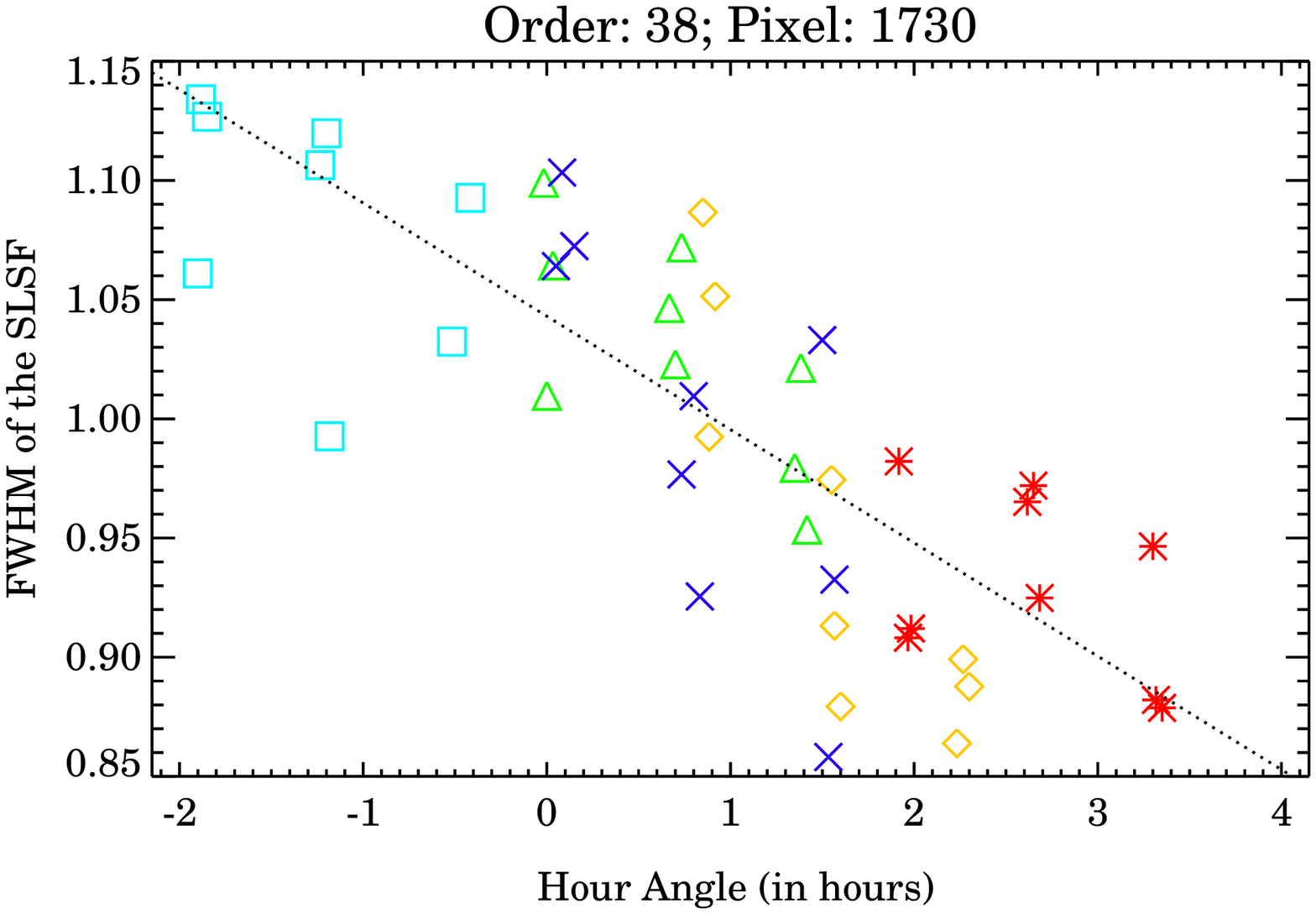}\\
\includegraphics[width=5cm]{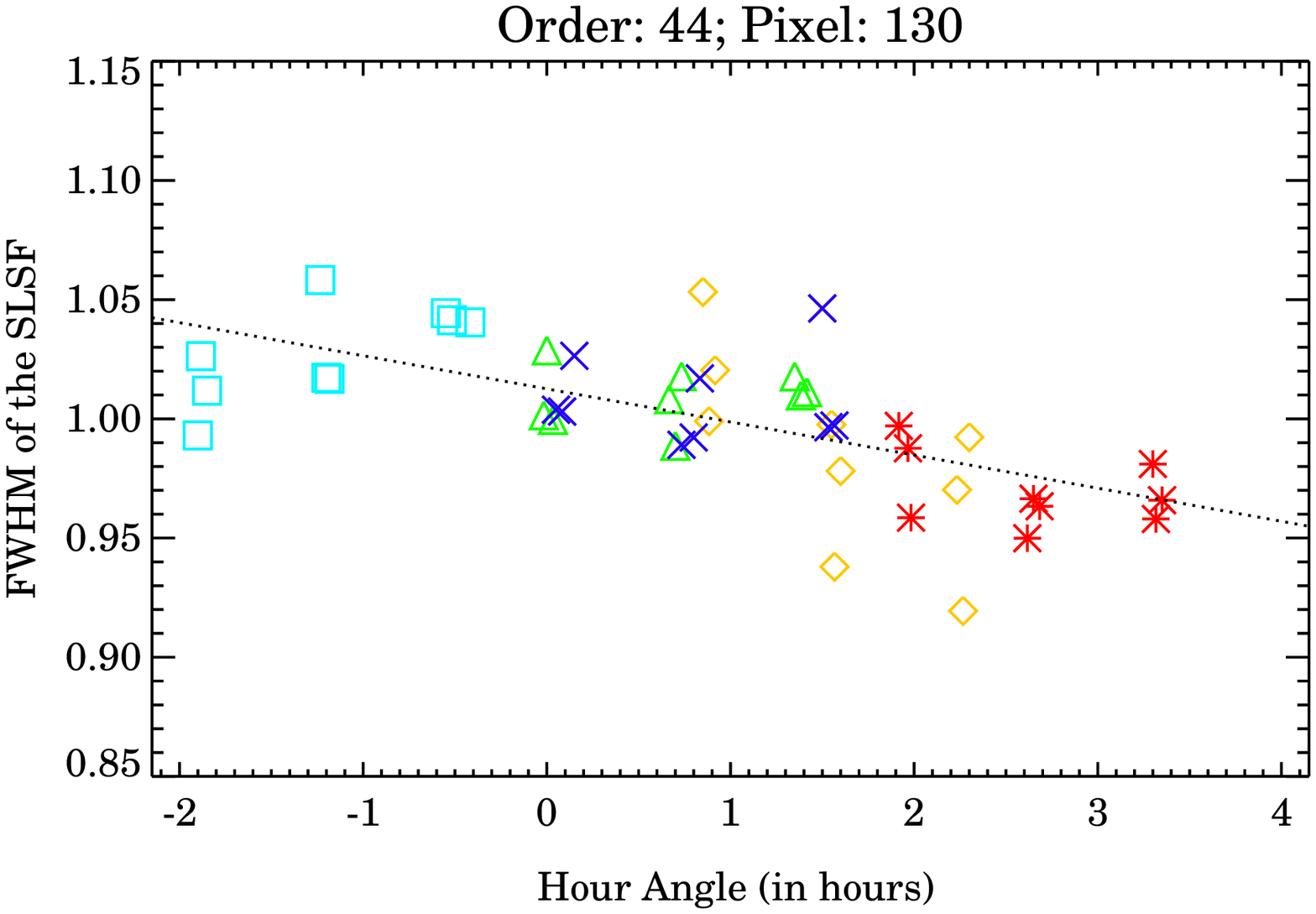}&\includegraphics[width=5cm]{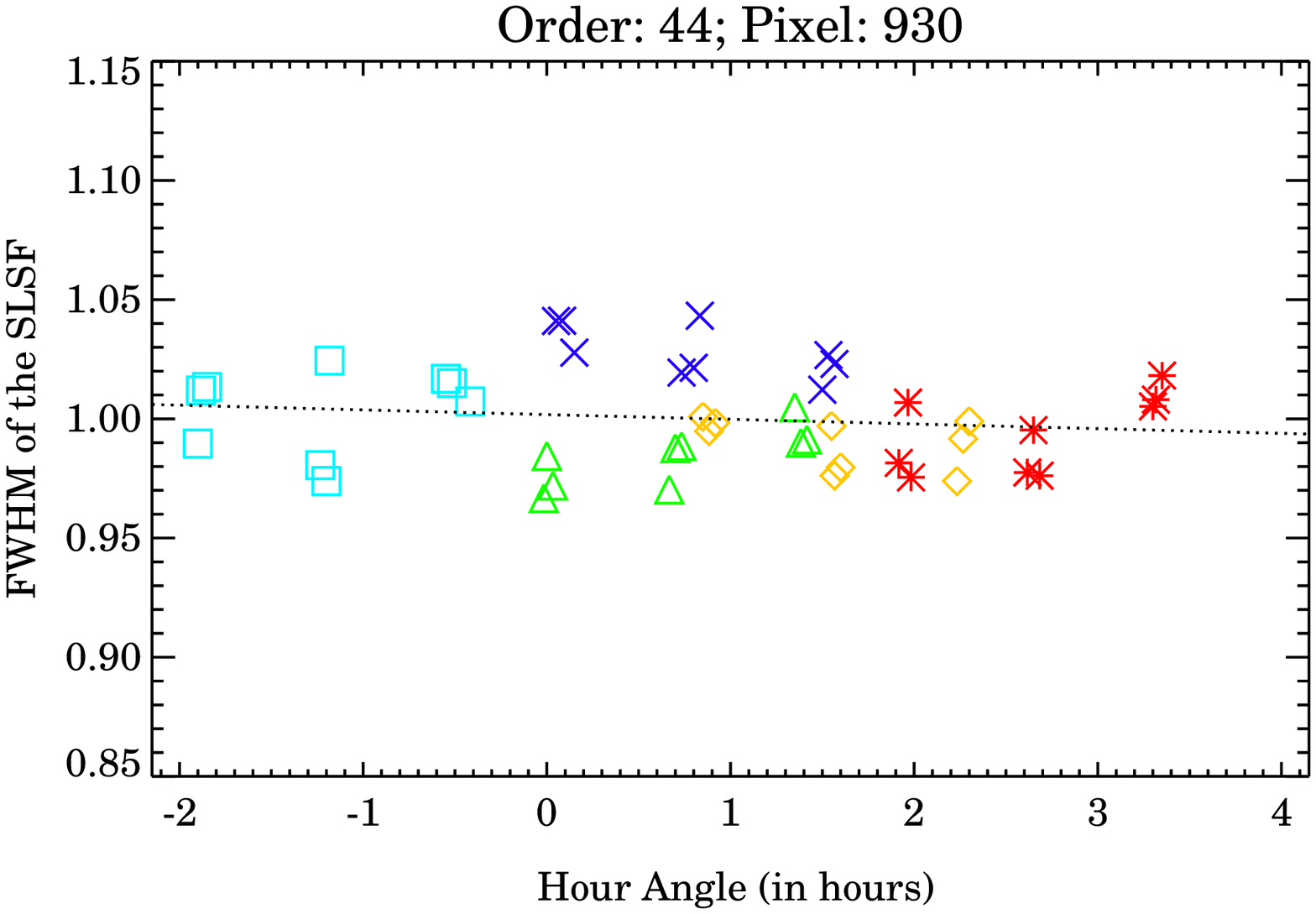}&\includegraphics[width=5cm]{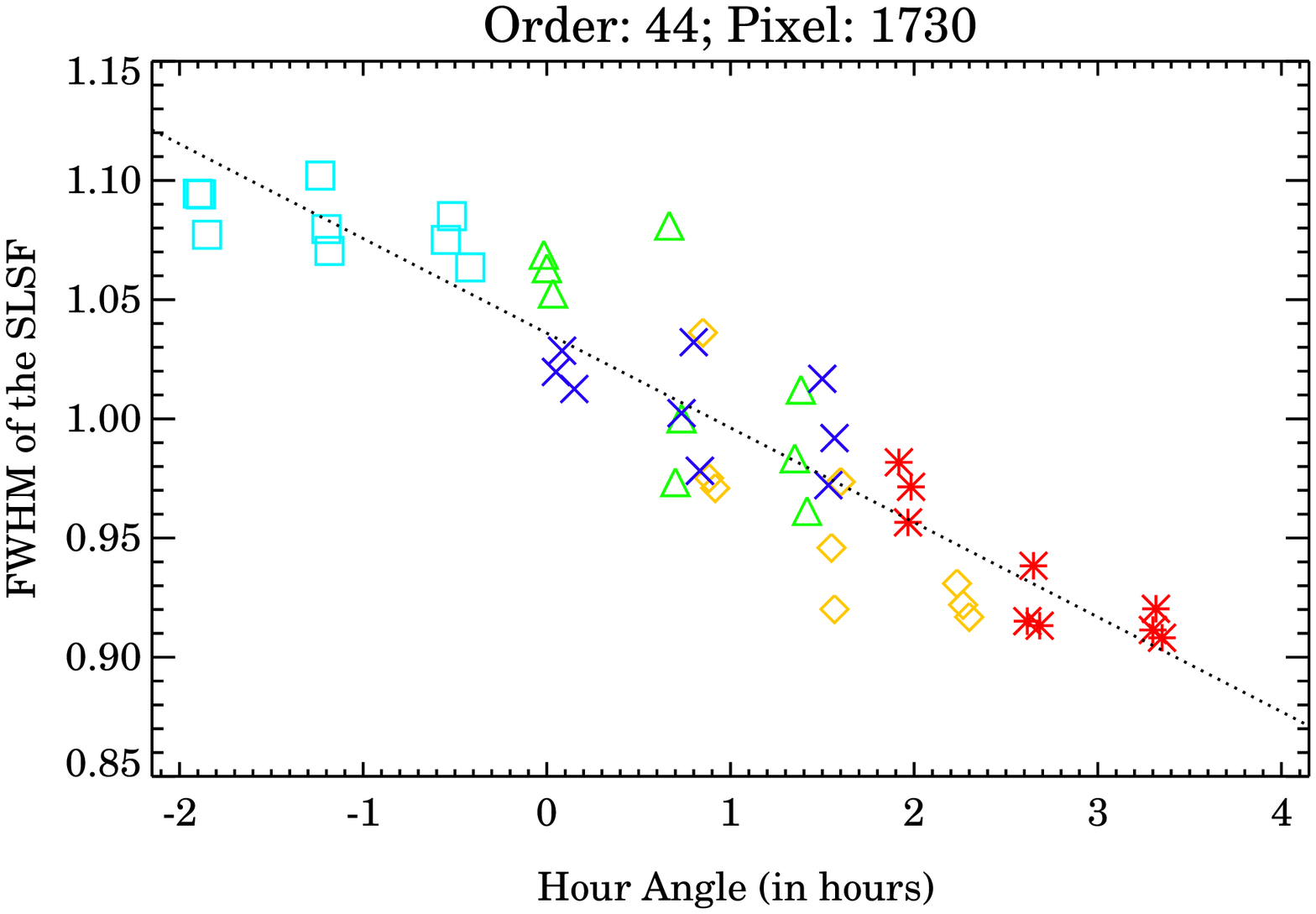}\\
\includegraphics[width=5cm]{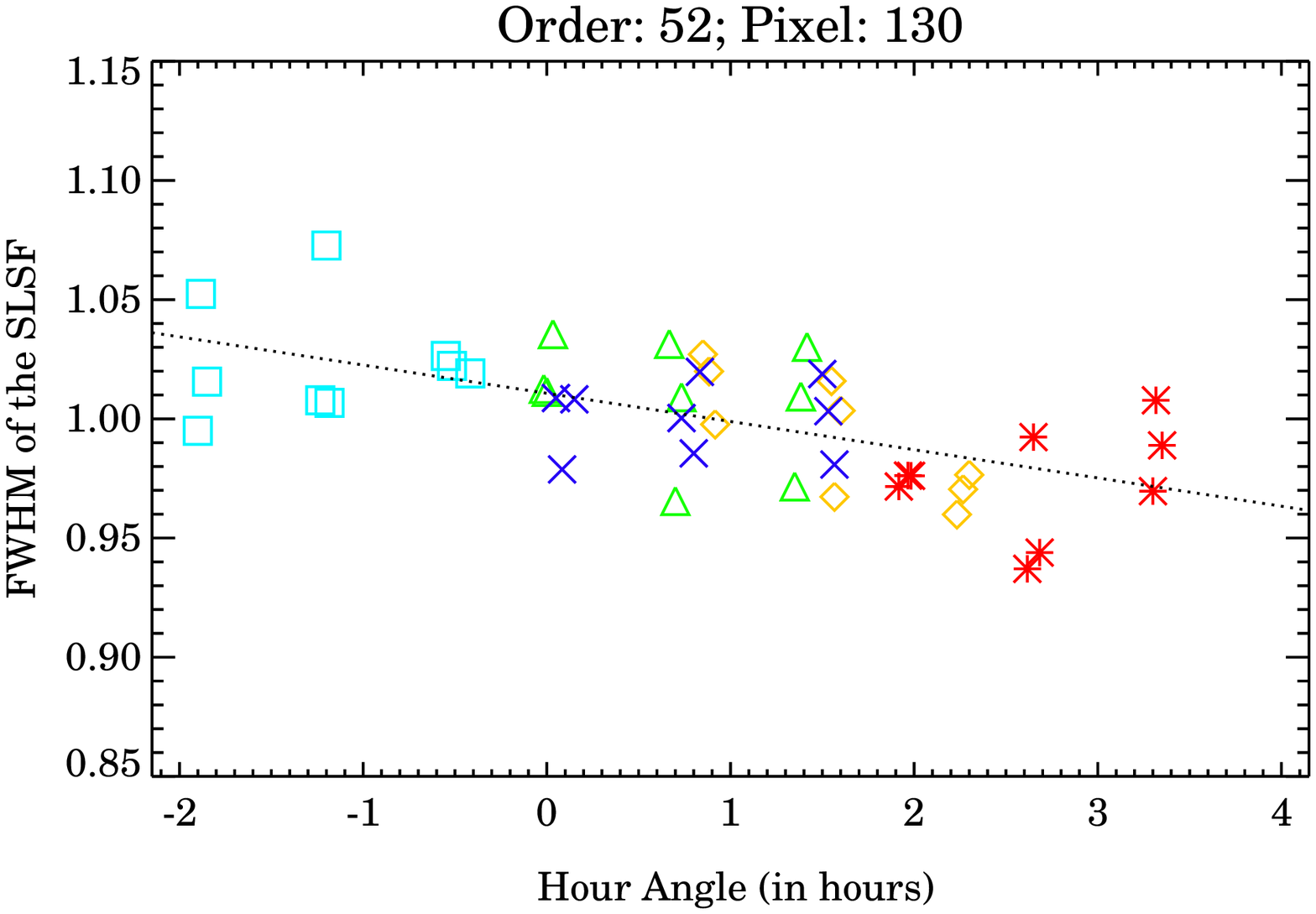}&\includegraphics[width=5cm]{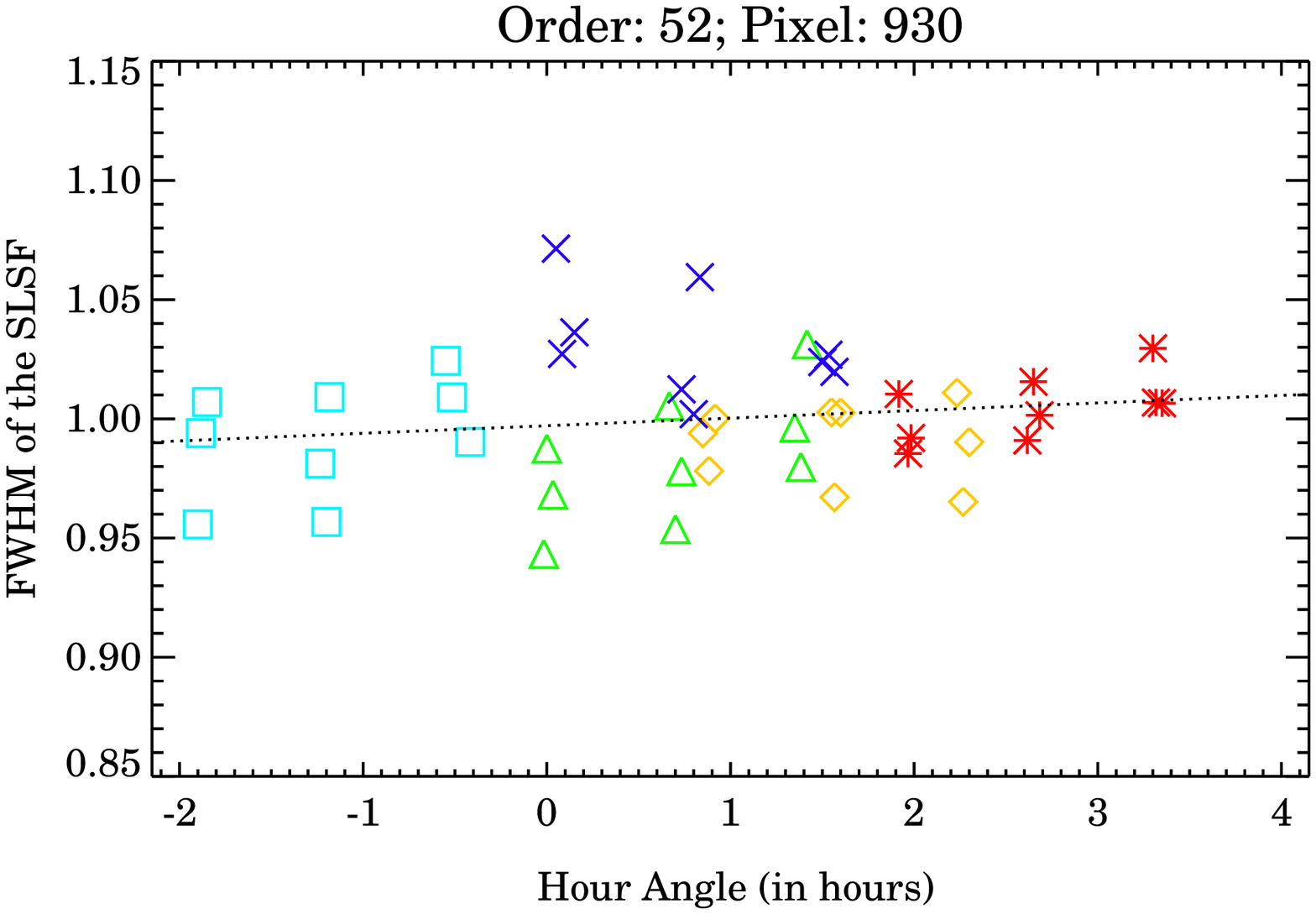}&\includegraphics[width=5cm]{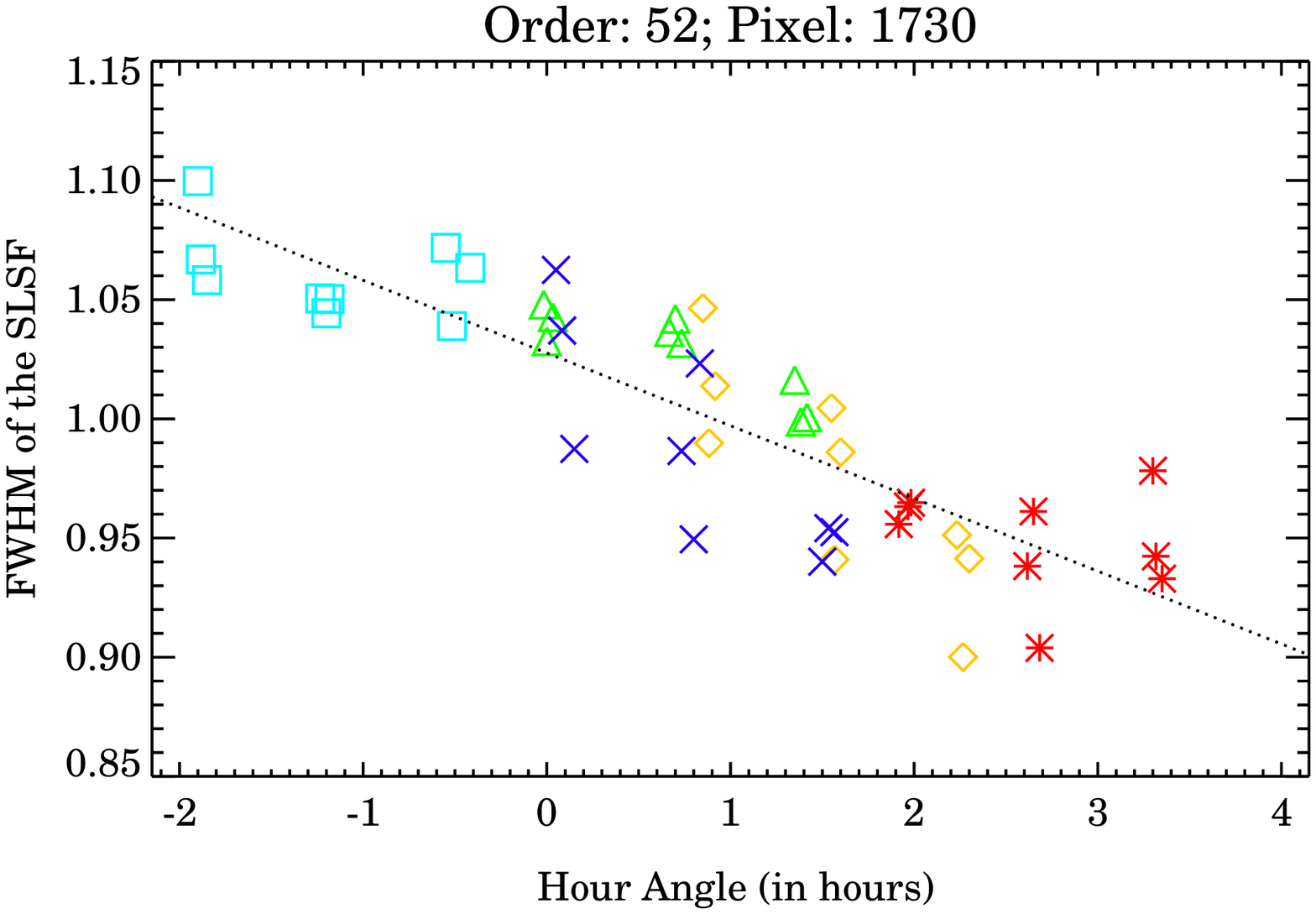}\\
	\end{tabular}
	\caption{FWHM of the SLSF as a function of hour angle for different parts of the CCD. Top-bottom is the cross-dispersion direction. Left-right is the dispersion direction.}
	\mylabelf{psfhaord}
\end{figure}

\section{Radial velocities}
\mylabels{rv}

In order to assess the improvement brought by the fiber feed, we have taken sets of observations with and without fiber and obtained velocities for these observations.
We considered three test stars (HD161797, HD143761 and HD166620) for which we respectively took 48, 25 and 15 fiber observations and 21, 15 and 42 slit observations. An iodine cell was used for wavelength calibration.
The RMS scatters of these velocities (after removing the linear trend from the binary orbit of HD161797 and after removing the signal from the planet in the HD143761 data) are listed in \tabl{vels}.
We also listed for each set of observations the mean signal-to-noise ratio (SNR). We see for HD161797 and HD143761 that the velocities are similar with and without fiber. However, the fiber velocities had a significantly lower SNR. Since RV scales with SNR, we would expect to get an improvement of about 30\% with the fiber feed when reaching the same SNR. This is confirmed by the radial velocities of HD166620 that show, with similar SNR with and without fiber, a 31\% improvement with the fiber scrambler.

\begin{table}[htbp]
	\centering
	\begin{tabular}{|c||c|c||c|c|} \hline
										& \multicolumn{2}{c||}{\textbf{With fiber}}	& \multicolumn{2}{c|}{\textbf{Without fiber}} \\ \hline
	{\textbf{Target}} & {\textbf{RMS}} & {\textbf{SNR}} & {\textbf{RMS}} & {\textbf{SNR}} \\ \hline
	HD161797 & 5.4~m.$\mbox{s}^{-1}$ & 113 & 4.9 m.$\mbox{s}^{-1}$ & 172 \\ \hline
	HD143761 & 9.2~m.$\mbox{s}^{-1}$ & 92 & 8.2 m.$\mbox{s}^{-1}$ & 144 \\ \hline
	HD166620 & 5.4~m.$\mbox{s}^{-1}$ & 93 & 8.4 m.$\mbox{s}^{-1}$ & 87 \\ \hline
	\end{tabular}
	\vspace{2mm}

	\caption{Radial velocities with and without fiber feed for three test stars: HD161797, HD143761 and HD166620.}
	\mylabelt{vels}
\end{table}

\section{Conclusions}
\mylabels{conclusions}

To summarize, in order to measure spectral line shifts smaller than one ten-thousandth of a pixel and stable for many months, we must reduce errors in our instrumental profile or SLSF, which cross-talks with our measurement of the Doppler shift. The SLSF variability can be broken down in coupling errors (slit or fiber illumination) and environmental instability. 

In this paper, we have shown improvement in SLSF stability using a fiber scrambler. These results demonstrate that coupling errors are the dominant source of instrumental profile variability at Lick observatory over short time scales. We show that double scrambler observations have a more stable SLSF than fiber observations by a factor~2. These have a more stable SLSF than slit observations by a factor~2. The double scrambler data still has residual RMS scatter. The source of this has not yet been identified but is likely to be modal noise, photon noise or modeling errors. We do not expect that the residual scatter can be caused by environmental effects due to the random nature of the variability. In the case of the Hamilton spectrograph at Lick, we have shown a strong correlation between SLSF variations and changes in star position in the sky, most likely due to changes of spectrograph illumination with varying telescope positioning.

Finally, we have shown a 30\% improvement in radial velocities for HD166620 between fiber and slit observations. For HD161797 and HD143761, the Doppler precision was slightly lower but at a significantly lower SNR. Scaled to the same SNR, fiber radial velocities should have an RMS scatter 30\% lower than the slit observations.

\acknowledgments
We acknowledge the support of the Planetary Society, who made possible the development and installation of the fiber feed at Lick observatory.
The authors would like to thank all staff at Lick observatory who have helped many times in the implementation of the fiber scramblers. We are also grateful to Kelsey Clubb for helping with the observing runs, Tom Blake for the iodine cell FTS spectra and Jeff Valenti for essential data normalization. A portion of this research was performed using EMSL, a national scientific user facility sponsored by the Department of Energy's Office of Biological and Environmental Research and located at Pacific Northwest National Laboratory. We also would like to thank the anonymous referee for helpful comments. DAF acknowledges research support from NSF grant AST-1036283 and NASA grant NNX08AF42G.

\end{document}